\documentclass[preprint, pra]{revtex4-1}

\usepackage[T1]{fontenc}
\usepackage{lmodern}
\usepackage{calc}
\usepackage{amsmath}
\usepackage{amssymb}
\usepackage{mathtools}

\usepackage[unicode=true]{hyperref}

\usepackage{graphicx}
\usepackage{dcolumn}
\usepackage{bm}
\usepackage{xcolor}
\usepackage{color}
\usepackage{float}

\begin{document}
\title{Spatial and Temporal Periodic Density Patterns in  Driven Bose-Einstein Condensates}

\author{A. del R\'{i}o-Lima}
\affiliation{Instituto de F\'{i}sica, Universidad Nacional Aut\'{o}noma de M\'{e}xico, C.P. 04510 Ciudad de M\'{e}xico, M\'{e}xico.}

\author{J. A. Seman}
\affiliation{Instituto de F\'{i}sica, Universidad Nacional Aut\'{o}noma de M\'{e}xico, C.P. 04510 Ciudad de M\'{e}xico, M\'{e}xico.}

\author{R. J\'{a}uregui}
\affiliation{Instituto de F\'{i}sica, Universidad Nacional Aut\'{o}noma de M\'{e}xico, C.P. 04510 Ciudad de M\'{e}xico, M\'{e}xico.}

\author{F. J.  Poveda-Cuevas}
 \affiliation{%
C\'atedras Conahcyt - Instituto de F\'{i}sica, Universidad Nacional Aut\'{o}noma de M\'{e}exico, C.P. 04510 Ciudad de M\'{e}xico, M\'{e}xico
}%



\date{\today}
\begin{abstract}

The study of collective excitations is a crucial tool for understanding many-body quantum systems. For instance, they play a central role in the exploration of superfluidity and other quantum macroscopic phenomena in Bose and Fermi systems. In this work we present a variational and a numerical study of a parametrically driven Bose-Einstein condensate confined in a cylindrical harmonic trap in which the aspect ratio can be varied from a prolate (cigar-shaped) to an oblate (pancake-shaped) system. The excitation can be applied by periodically modulating the harmonic frequencies of the trap or, alternatively, the interatomic interaction strength at a frequency that matches that of the system breathing mode. As a result, we observe the formation of dynamical density patterns that depend on the geometry of the trap: a fringe pattern in a prolate system and a ring pattern in an oblate one. By decomposing the total energy into its kinetic, potential, and interaction terms, we show that the onset of these patterns coincides with the redistribution of kinetic energy along the weakly trapped directions of the sample, indicating the three-dimensional nature of the studied phenomena. Finally, our analysis shows that the difference between the two excitation mechanisms lies on the system stability. Modulating the trap destabilizes the system quicker than modulating the interactions, leading to earlier formation of the patterns.

\end{abstract}

\maketitle 


\section{Introduction}

Pattern formation is ubiquitous and present on different scales in nature. It can be encountered in biological and chemical systems, condensed matter physics, hydrodynamics, nonlinear optics, and other systems \cite{Philip-book2001}. In biology, its study is of special relevance because the spontaneous emergence of patterns in time and space. This phenomena known as self-organization is a manifestation of collective interactions that are responsible for life \cite{Turing1952}. In chemistry, systems undergoing chemical reactions show many different pattern-forming phenomena that combine thermodynamics with reaction-diffusion dynamics \cite{Lindgren2024}. In physics, the study of pattern formation has been present in understanding non-equilibrium processes accompanying, e.g., phase transitions and in the theory of non-linear dynamics. Many mechanical systems change from simple to complex behavior in response to modifications in their control parameters \cite{Gollub1999}. Non-linear dynamics theory allows a better comprehension  of biological and chemical systems by simulating them with analogue physical systems with a simpler structure \cite{Cross-rmp65}.

Nowadays, the concept of instability is at the center of understanding pattern formation. Spatial and temporal patterns emerge when a uniform system is driven into unstable states which will deform by large amounts in response to infinitesimally small perturbations.  Perhaps the best known instabilities in fluid dynamics are Rayleigh-B\'enard convection \cite{Getling-book1998}, Taylor-Couette flow \cite{Taylor-ptrsla223}, and  parametric excitations \cite{Champneys-book2009}. The last type has been extensively studied in classical fluids subject to a single parametric excitation mode; under a linear approximation they can be described by the well-known Mathieu equation \cite{McLachlan-book1947, Arscott-book2014}. Nevertheless, out of the linear regime, its treatment increases in complexity due to the hydrodynamic nature, and it remains unclear how they evolve into chaotic behavior and turbulence. 

The achievement of the experimental generation of quantum fluids and its theoretical description has allowed a new path for  understanding pattern formation, with the advantage of fine control under the system. In classical systems, the container's geometry imposes spatial boundary conditions, and the fluid's viscosity affects the formation of the patterns \cite{Miles-annrevfluidmech22, Kumar-proca452, Kumar-pre62, Cerda-prl78, Barrio-pre56}.
For quantum fluids constituted by cold atomic gases, the geometry is controlled via an external optical or magneto-optical trapping potential \cite{Guo-naturephys20}. Besides, the conservative interaction strength between the atoms can be modified via Feshbach resonances \cite{Chin-revmodphys82, Schunck-pra71, Hulet-revsciinstr91}.
Control parameters allow for regulating the coherence properties of the system, along with the characteristics of propagation and geometry of the excitations. This opens the possibility of studying the similitudes and differences between classical and quantum systems and simultaneously, learning about the dynamics of quantum systems under specific local and global conditions. 
One way to create patterns in a quantum fluid is through the system's collective excitations influenced by the interaction between particles. 
For harmonically trapped quantum gases the quasi stationary collective modes  of a  sample that changes  its  position,  angle,  or  form  without  undergoing  significant  volume changes
are classified as a surface modes. Excitations of this kind have been used to study the transition
from hydrodynamic to collisionless behavior in  ultracold atomic gases \cite{Stamper1998, Buggle2005, Wright2007}.
If the oscillation involves significant changes of the volume, and  as a consequence of the density of the fluid, the mode is classified
as a compression or breathing mode \cite{Bartenstein-prl92, Tey-prl110, Huang-pra99}. 

In the last years, various types of parametric modulations have been applied to different quantum fluids and their density patterns have been studied \cite{Kronjaeger-prl105, Zhang-natphys16, Staliunas-prl89, Nicolin-pra76, Staliunas-pra70, Brito2023, Maity2020}. For example, non-linear parametric excitations known as Faraday waves were observed, for the first time, in a weakly interacting Bose-Einstein Condensate of $^{87}\mathrm{Rb}$ in a cigar-shaped trap \cite{Engels-prl98}, and in a $^{23}\mathrm{Na}$ BEC \cite{Smits-prl121}. These patterns were demonstrated to evolve to a granulation regime in a $^{7}\mathrm{Li}$ BEC \cite{Nguyen-prx9}. Furthermore, density patterns have been generated in a strongly interacting superfluid of $^6\mathrm{Li}$ by a parametric modulation of the radial trap frequency \cite{HernandezRajkov-njp23}. Due to the versatility and high control over superfluids, different types of patterns can be generated depending on how the system is excited. K. Fujii et al. demonstrated that suppressing reflections at the boundaries can reproduce dynamics similar to an infinitely extended system. Thus, they report the spontaneous emergence of a square lattice state in a two-dimensional BEC with absorptive boundary conditions  \cite{Fujii2024, Oberthaler2023}. In the work of K. Kwon et al. \cite{Kwon2021}, star-shaped patterns are formed in a pancake-shaped BEC when they modulate the scattering length with frequencies related to the radial confining trap frequency.  Within all the possibilities, a particular type of pattern arises when the system is modulated exactly with the breathing mode frequency of the atomic cloud, and thus, the system oscillates in resonance. These spatial and temporal patterns over the quantum fluid are commonly called Faraday Waves due to their analogy with the classical system \cite{Michael1831}, in which a wave is created in the perpendicular plane to the direction in which the modulation was introduced, and the pattern's frequency is equal to half the frequency of the parametric modulation \cite{Miles-annrevfluidmech22}. Therefore, the trap geometry, the type of parametric modulation, and the frequency at which the system is modulated affect the development and stability of emergent patterns.


This work aims to give insight into the pattern formation mechanism in a pancake-shaped and cigar-shaped trap when the axial or radial trap frequency, respectively,  or the scattering length are modulated with the breathing mode frequency. A variational analysis is used to obtain the frequency of the collective mode, which exhibits an explicit dependence on the geometry. Numerical simulations of the evolution of the Gross-Pitaevskii equation are carried out under a parametric modulation beyond the simplified dimensional reduction analysis. By means of a fidelity function, we study the deviation from a given density pattern. For this, we choose the wavefunction evaluated at the time in which the density patterns arise as the initial state and compare it with wavefunctions at longer times.

Although our system shares some characteristics with the Faraday Waves excitations, we will name them \textit{density patterns} instead. We want to emphasize that our patterns evolve while the system oscillates in resonance with a collective excitation. Consequently, we have a system with an amplitude of oscillation that in the simplest description would diverge and in practice is unstable. 
Close to resonance, small changes in the frequency modify significantly the behavior of the system. Consequently, it is very important to characterize the system response in the resonance frequency region. It will be shown that the spatial and temporal periodic modulation of the density in the resonance pattern is due to the interference of the excitations generated by the parametric modulations. The interference of these excitations with opposite velocities creates the spatial pattern. The temporal periodic pattern presents a frequency equal to the modulation frequency. In contrast with a cigar-shaped trap, in the oblate trap, the excitations propagate in a radial and axial direction, creating new types of patterns. Our oblate BEC simulations present ring patterns in the $xy$ plane, contrasting with the square patterns observed in the infinitely extended limit \cite{Fujii2024, Oberthaler2023}.  The simulations for the prolate system reproduced the fringe patterns observed previously experimentally \cite{Engels-prl98, Smits-prl121, Nguyen-prx9, HernandezRajkov-njp23}. Finally, we found a difference in the patterns' dynamic due to the anisotropy (isotropy) of the modulation of the trap frequency (interaction), which has not been reported earlier and results to be important in the stability of the oblate trap. 

The article is structured in the following way. In Section \ref{sec:parametricexcitation} we discuss how to obtain the collective modes of the BEC through a variational method. In Section \ref{sec:numericalsimulation}, the details of the numerical simulation are discussed together with the parameters that characterize the system. Section \ref{sec:solutions} presents the results and discussion of the numerical simulations. Finally, in Section \ref{sec:Conclusions} the conclusions and perspectives are given.

\section{Parametric excitation of a BEC}
\label{sec:parametricexcitation}
We are interested on the dynamics of interacting Bose-Einstein condensates under  parametric periodic excitations  with a frequency  equal or close enough to that of the breathing mode.
In the mean field approximation at $T=0$, the system is described by the  Gross-Pitaevskii equation (GPE), 
\begin{equation}
(\imath-\gamma_d)\hbar \frac{\partial}{\partial t}\Psi (\mathbf{r},t)=\left( -\frac{\hbar^2}{2M}\nabla^2 + V_{\mathrm{ext}}(\mathbf{r},t)+\frac{4\pi \hbar^2 a_s(t)}{M}|\Psi (\mathbf{r},t)|^2\right)\Psi (\mathbf{r},t). \label{GPE}
\end{equation}
 We introduce a temporal dependence in both, the external (trapping) potential $V_{\mathrm{ext}}(\mathbf{r},t)$ and in the interacting parameter $a_s(t)$, allowing two different forms of parametric modulation of the BEC density, i.e., by the modulation of the interaction between particles through the scattering length, or alternatively, by the modulation of the radial or axial trap frequency for the prolate or oblate geometry, respectively. We do not consider the case in which both quantities, $V_{ext}$ and $a_s$, are simultaneously modulated; the system is excited by modulating only one of them. A dimensionless phenomenological parameter, $\gamma_d$, is included in \eqref{GPE} to account for dissipation mechanisms. An important source of dissipation is, for instance, the presence of a thermal component surrounding the BEC which can damp excitations in the system \cite{Choi1998}. We set $\gamma_d=0.01$ since this value reproduced well the experimental observations of the generation of FW in a cigar-shaped BEC \cite{HernandezRajkov-njp23}. 
  
Mean field equation \eqref{GPE} can also be used to describe dilute Fermi gases in the degenerate BEC limit. In general, the interactions in a dilute Fermi gas are parametrized by the dimensionless interaction strength $\eta \equiv (k_F a_s)^{-1}$, where $k_F$ is the Fermi wavevector, $k_F = \sqrt{2mE_F} / \hbar$, $E_F$ is the Fermi energy, $m$ is the atomic mass, and $a_s$ is the interatomic scattering length. The system is in the strongly interacting regime if $-1\leq \eta \leq 1$ the so called BEC-BCS crossover, while $\eta > 1$ and $\eta < -1$ correspond, respectively, to the BEC and BCS limits. To equation \eqref{GPE} be appropriate to describe the BEC limit, the condition $\eta >1$ requires to be satisfied. 
   
We consider an external harmonic potential in cylindrical coordinates $\{r,\theta,z\}$, 
\begin{equation}
V_{\mathrm{ext}}\left(\mathbf{r},t\right)=\frac{1}{2}m(r^2 \omega_r^2(t)+z^2 \omega_z^2(t))
\label{eq:potential_mod}
\end{equation}
where the temporal dependence of $\omega_r(t)$ and $\omega_z(t)$ indicates the possibility of  modulating any of them. It is important to clarify that the driving is introduced only through the most confining frequency. Hence, in the case of a prolate (also denoted as "cigar-shaped") BEC, $\omega_r$ is the frequency that we modulate while keeping $\omega_z$ constant. Equivalently, for an oblate (or "pancake-shaped") system, $\omega_z$ is modulated while $\omega_r$ is kept unchanged.  Also, as previously explained, in this case in which the trap is driven, the scattering length remains constant during the whole dynamics of the system.

In current experiments, optical dipole traps are widely employed to generate the harmonic potential in which the BEC is confined \cite{Grimm-advamo42}. These traps are produced by focusing a far red-detuned laser beam. The harmonic frequencies are related to the power of said beam, $P$, as $\omega_k \propto \sqrt{P}$. Therefore, the modulation of the trap frequencies can be achieved by oscillating the power of the optical beam confining the atoms. Since the modulation can be considered as a small perturbation, the trap frequencies are correctly described as 
\begin{equation}
 \omega_k(t)=\omega_{k0} \sqrt{1+\alpha \cos \Omega (t-t_0)}\approx \omega_{k0} \left(1+\tfrac{\alpha}{2}\cos \Omega (t-t_0) \right), \quad  k=r,z,
 \end{equation}
where  $\omega_{k0}$ is the initial value of the trap frequency that is modulated, $\alpha$ is the amplitude of the modulation satisfying $\alpha \ll 1$, $\Omega$ is the external frequency of modulation, and $t_0= -\pi n/(2\Omega)$ with $n$ as an integer is a phase  that ensures that a $t=0$, the experimental modulation is zero. 

Similarly, in the case in which the system is excited by modulating the scattering length, $a_s(t)$ (while keeping the trapping potential constant), we consider
\begin{equation}
g\left(t\right) = \frac{4\pi \hbar a_s(t)}{M}, \quad  a_s\left(t\right) = a_{s0} \left(1+\alpha \cos \Omega \left(t-t_0\right) \right),\label{eq:gt}
\end{equation}
where $a_{s0}$ is the initial scattering lenght.
 
As we describe in the following Subsection \ref{subsec:VariationalApproach}, there is a connection between the linearized equations of motion of the system, and  the linearized dynamics of classical parametric oscillators in terms of solutions of the Mathieu equation, which in its canonical form reads
\begin{equation}
\frac{d^2 h}{d\eta^2} + [\beta - 2 \kappa\cos(2\eta)]h = 0. \label{mathieueq}
\end{equation}
If the modulation of the trapping frequency is in the direction $k_0$, we identify $h$ as the width of the Gaussian distribution in that direction   $\sigma_{k_0}$, and the adimensional terms $\eta,\, \beta$ and $\kappa$ are then given by
\begin{equation}
\eta = \frac{\Omega (t-t_0)}{2}, \quad\beta = \Big(\frac{2\omega_{k0}}{\Omega}\Big)^2, \quad \kappa = \alpha^2\Big(\frac{2\omega_{k0}}{\Omega}\Big)^2
\end{equation}

It is well known \cite{landau1982mechanics} that depending on the values of $\beta$ and $\kappa$, the solutions are stable or unstable.
 For $\kappa\ll \beta$ a Floquet  analysis of these  solutions  results in the identification of
 quasi-periodic harmonic oscillations at secular frequencies
$$\Omega_n/2 \sim \omega_{k0}/n$$ where $n$ is an integer that enumerates them. This results from taking into account that (i)  the general expression for the even $\mathrm{ce}_n(\eta)$ and odd $\mathrm{se}_n(\eta)$ periodic solutions of the Mathieu equation can be written as a series of the $\cos r\eta$ and $\sin r\eta$ functions with $r$ and $s$ integer numbers, and (ii) for
$\kappa=0$, $\beta_n =n^2$ corresponds to an eigenvalue of the resulting equation. The frequencies $\Omega_n$ define the classical- linear-resonance condition.  The width of these resonances can also be estimated for $\kappa\ll \beta$ yielding $\Delta_n = \kappa^n/[2^{2n-3}((n-1)!)^{1/2}]$ \cite{Bell57}.
The secular oscillations, in general, are  superimposed  with the so-called micromotion
driven by the time-dependent potential. It has been shown that the quantum analogue of the classical description of parametric periodic excitations yields similar results \cite{Jauregui2001} in terms of expectation values and correlations of the relevant quantum operators.

\subsection{Variational Approach} \label{subsec:VariationalApproach}

The GPE can be equivalently described as a variational equation  derived from the non stationary Lagrangian density \cite{Perez-Garcia-pra56}, 
\begin{equation}
\mathcal{L}\left(\mathbf{r},t\right)=\frac{\imath\hbar}{2}\left(\Psi^{*}\frac{\partial\Psi}{\partial t}-\Psi\frac{\partial\Psi^{*}}{\partial t}\right)-\frac{\hbar^{2}}{2m}\left|\nabla\Psi\right|^{2}-V_{\mathrm{ext}}\left(\mathbf{r},t\right)\left|\Psi\right|^{2}-\frac{1}{2}g\left(t\right)\left|\Psi\right|^{4}.\label{lagrangiandensity}
\end{equation}
 
The variational approach to the non-linear dynamics is based here on a Gaussian {\it ansatz} for the condensate wavefunction
\begin{equation}
    \Psi\left({\bf r};\{\bm{q}\left(t\right)\}\right)=f(t) \exp\left[-\frac{r^{2}}{2\sigma_{r}^{2}(t)}+\imath\beta_{r}(t)r^{2}\right]\exp\left[-\frac{z^{2}}{2\sigma_{z}^{2}(t)}+\imath\beta_{z}(t)z^{2}\right],
\label{GaussianAnsatz}
\end{equation}
with time dependent radial $\sigma_r$ and axial $\sigma_z$  widths, as well as the radial $\beta_r$ and axial $\beta_z$ phases taken as variational parameters. They are compactly denoted by $\{\bm{q}(t)\}=\{\sigma_r, \sigma_z,  \beta_r, \beta_z\}$. The function $f(t)$ 
\[
f(t)=\sqrt{\frac{N}{\pi^{3/2}\sigma_{z}\left(t\right)\sigma_{r}^{2}\left(t\right)}},
\] 
guarantees both the accomplishment of the continuity equation and the normalization of the ansatz function to a constant atom number $N$. Our ansatz does not include the possibility of a vortex at the center, which could be incorporated through a  factor of the form $\left(r \mathrm{e}^{ \imath \theta }\right)^{\ell}$.

The non-linear character of the density dependent interaction couples the
oscillations in different directions \cite{Nicolin-physa389}.
The characterization of  the breathing mode for different trap geometries is a relevant step for the proper identification of  the resonant regime.
When the system is modulated by a frequency that equals the one of the breathing mode, 
patterns are formed in the plane perpendicular to the one where the modulation was introduced \cite{Jackson-prl88, Perez-Garcia-prl77}.
In this Section,  we perform a variational approach to obtain the approximate frequencies that would lead to resonant effects \cite{Garcia-Ripoll-prl83}.

Substituting the {\it ansatz} of Eq.~\eqref{GaussianAnsatz} into Eq.~\eqref{lagrangiandensity} and integrating over the spatial coordinates gives a set of equations for the variational parameters $\sigma_r$, $\sigma_z$, $\beta_r$, and $\beta_z$ describing the optimal representation of $\Psi$ in terms of Gaussian functions.
After some algebra, the dynamic is reduced to two coupled equations for the Gaussian widths $\sigma_r$ and $\sigma_z$ since the equations for phases  $\beta_r$ and $\beta_z$ are in terms of these Gaussian widths.

For simplicity, the following variational equations of motion for the Gaussian widths are presented in their dimensionless form. For the longitude variables, their dimensionless form is defined in units of the natural harmonic oscillator length
 \begin{equation}
 l_0=\sqrt{\frac{\hbar}{m\omega_0}}\label{eq:holength}
 \end{equation}
and the time takes units of the inverse of the natural frequency $\omega_0$, which is equal to $\omega_{r0}$ for the prolate trap and $\omega_{z0}$ for the oblate trap.   

Depending on the geometry of the external trap and the parametric modulation, three relevant cases are identified,
\begin{enumerate}
    \item Trap modulation in a prolate ($\omega_{r0} \gg \omega_{z0}$) geometry. Only the radial trap frequency $\omega_r$ is modulated, keeping $\omega_z$ and $g$ constant,
    \begin{subequations}
\begin{eqnarray}
\ddot{\sigma}_{r}+\omega_{r0}^{2}(1+\alpha \cos \Omega (t-t_0))\sigma_{r}&=&\frac{1}{\sigma_{r}^{3}}+\sqrt{\frac{2}{\pi}}\frac{N a_{s}}{\sigma_{r}^{3}\sigma_{z}},\label{sigmar} \\
\ddot{\sigma}_{z}+\omega_{z0}^{2}\sigma_{z}&=&\frac{1}{\sigma_{z}^{3}}+\sqrt{\frac{2}{\pi}}\frac{N a_{s}}{\sigma_{r}^{2}\sigma_{z}^{2}} \label{sigmaz},
\end{eqnarray} \label{trap_modulation_prolate}
\end{subequations}
    \item Trap modulation in an oblate ($\omega_{r0} \ll \omega_{z0}$) geometry. Only the axial trap frequency $\omega_z$ is modulated, keeping $\omega_r$ and $g$ constant,
    \begin{subequations}
\begin{eqnarray}
\ddot{\sigma}_{r}+\omega_{r0}^{2}\sigma_{r}&=&\frac{1}{\sigma_{r}^{3}}+\sqrt{\frac{2}{\pi}}\frac{N a_{s}}{\sigma_{r}^{3}\sigma_{z}}, \label{sigmar} \\
\ddot{\sigma}_{z}+\omega_{z0}^2(1+\alpha \cos \Omega (t-t_0))\sigma_{z}&=&\frac{1}{\sigma_{z}^{3}}+\sqrt{\frac{2}{\pi}}\frac{N a_{s}}{\sigma_{r}^{2}\sigma_{z}^{2}} \label{sigmaz},
\end{eqnarray} \label{trap_modulation_oblate}
\end{subequations}
    \item Scattering length modulation in both geometries. The radial $\omega_r$ and axial $\omega_z$ trap frequencies are kept constant,
    \begin{subequations}
\begin{eqnarray}
\ddot{\sigma}_{r}+\omega_{r}^{2}\sigma_{r}&=&\frac{1}{\sigma_{r}^{3}}+\sqrt{\frac{2}{\pi}}\frac{N a_{s0}(1+\alpha \cos \Omega (t-t_0))}{\sigma_{r}^{3}\sigma_{z}} \label{sigmar} \\
\ddot{\sigma}_{z}+\omega_{z}^{2}\sigma_{z}&=&\frac{1}{\sigma_{z}^{3}}+\sqrt{\frac{2}{\pi}}\frac{N a_{s0}(1+\alpha \cos \Omega (t-t_0))}{\sigma_{r}^{2}\sigma_{z}^{2}} \label{sigmaz}.
\end{eqnarray} \label{interaction_modulation}
\end{subequations}
\end{enumerate}
The excitation frequencies can be scaled in  terms of the highest trapping frequency, i. e.,  $\Omega =: f \cdot \omega_{r0}$ for the prolate trap and $\Omega = : f \cdot \omega_{z0}$ for the oblate trap, where $f$ is a proportional factor. The variational  result, presented in this Section, shows that in the prolate (oblate) trap, the factor $f$ for resonance is $f=2.0$ ($f=1.80)$ for the most elongated cases. Similar results were obtained in \cite{Stringari1996} by solving the hydrodynamic equations for a dilute gas of bosons.

Although most of the work made in resonance patterns in oblate traps considers the system to be in 2D \cite{Combescot-pra74}, we remark that the geometry of the system of interest in this work  does not necessarily satisfy the established condition for a dimensional reduction, meaning that our system is always tridimensional. In such a case, the coupling dynamics in the axial and radial directions will be used to obtain more information  about the pattern formation in the resonance regime.

Explicit solutions to equations \eqref{trap_modulation_prolate}-\eqref{interaction_modulation} can be obtained by considering a linear response in the change of the Gaussian widths,  
  \begin{equation*}
\sigma_r(t)=\sigma_{r0}+\delta_r(t),\quad  \sigma_z(t)=\sigma_{z0}+\delta_z(t). 
  \end{equation*}
 
Thus, the initial Gaussian widths $\sigma_{r0}$ and $\sigma_{z0}$ only change in time by a small amount $\delta_r(t)$ and $\delta_z(t)$, respectively, that satisfy the inequality  $\delta_j\ll\sigma_{j0}$ (with $j=r,z$). With these linear considerations, we arrive to the following results: 

\begin{enumerate}
 \item Prolate geometry with a modulation in the radial trap frequency.
 \begin{subequations}
\begin{eqnarray}
\ddot{\delta}_{r}+\left[A_{r}-2Q_{r}\cos \Omega (t-t_0)\right]\delta_{r}&=&-G\delta_{z}+f_{r0}\cos \Omega (t-t_0)\\
\ddot{\delta}_{z}+A_{z}\delta_{z}&=&-2G \delta_{r}.
\end{eqnarray} \label{delta_prolate}
\end{subequations}
\item Oblate geometry with a modulation in the axial trap frequency.
     \begin{subequations} 
     \begin{eqnarray}
\ddot{\delta}_{r}+A_{r}^{\prime}\delta_{r}&=&-G\delta_{z} \\ 
 \ddot{\delta}_{z}+\left[ A_{z}^{\prime}-2Q_{z}^{\prime}\cos\Omega (t-t_0)\right] \delta_{z}&=&-2G\delta_{r}+f_{z0}^{\prime}\cos\Omega (t-t_0).
     \end{eqnarray} \label{delta_oblate}
 \end{subequations}
\item Oblate and prolate geometry with a modulation in the scattering length. 
\begin{subequations}
    \begin{eqnarray}
\ddot{\delta}_{r}+\left[A_{r}^{\prime\prime}-2Q_{r}^{\prime\prime}\cos \Omega (t-t_0)\right]\delta_{r}&=&\left[G+b\cos \Omega (t-t_0)\right]\delta_{z}+f_{r0}^{\prime\prime}\cos \Omega (t-t_0) \\
\ddot{\delta}_{z}+\left[A_{z}^{\prime\prime}-2Q_{z}^{\prime\prime}\cos \Omega (t-t_0)\right]\delta_{z}&=&-2\left[G+b\cos \Omega (t-t_0)\right]\delta_{r}+f_{z0}^{\prime\prime}\cos \Omega (t-t_0). 
    \end{eqnarray} \label{delta_as}
\end{subequations}
\end{enumerate}
The definition of the common parameter $\mathrm{g}$, in the three sets of coupled equations,  is given by, 
\begin{equation*}
    G=\sqrt{\frac{2}{\pi}}\frac{Na_{s}}{\sigma_{r0}^{3}\sigma_{z0}^{2}}.
\end{equation*}
The other ones, describing the dynamics of each case, are given in Table \ref{table:1}. Note that these sets of linearized equations have the form of a driven Mathieu equation. Thus, the connection between the linearized equations of motion of the system and the linearized dynamics of classical parametric oscillators is given by driven Mathieu equations. To characterize the driven Mathieu equations \eqref{delta_prolate}-\eqref{delta_as}, we can use Eq. \eqref{mathieueq} and applied some of its properties, as its resonance width given by $\Delta_n$ (Section \ref{sec:parametricexcitation}).

When no external modulation is considered, $\Omega=0$,  the three cases are described by the same dynamical equation, which can be written in the  matrix equation form, 
\begin{equation}
\ddot{\bm{\delta}}(t)=\mathcal{W}_{0} \, \bm{\delta}(t)  \label{matrix_eq}
\end{equation}
where $\bm{\delta}$ is the vector with the Gaussian width variations as its components, $\bm{\delta}(t)=( \delta_r,\delta_z)^{\dagger}$, and $\mathcal{W}_{0}$ is the natural matrix defined as
\begin{equation}
\mathcal{W}_{0}=\left(\begin{array}{cc}
4\omega_{r0}^{2} & \sqrt{\frac{2}{\pi}}\frac{N a_{s}}{\sigma_{r_{0}}^{3}\sigma_{z_{0}}^{2}}\\
2\sqrt{\frac{2}{\pi}}\frac{N a_{s}}{\sigma_{r_{0}}^{3}\sigma_{z_{0}}^{2}} & 3\omega_{z0}^{2}+\frac{1}{\sigma_{z_{0}}^{4}}
\end{array}\right).
\end{equation}

\begin{figure}[h]
    \includegraphics[width=0.8\textwidth]{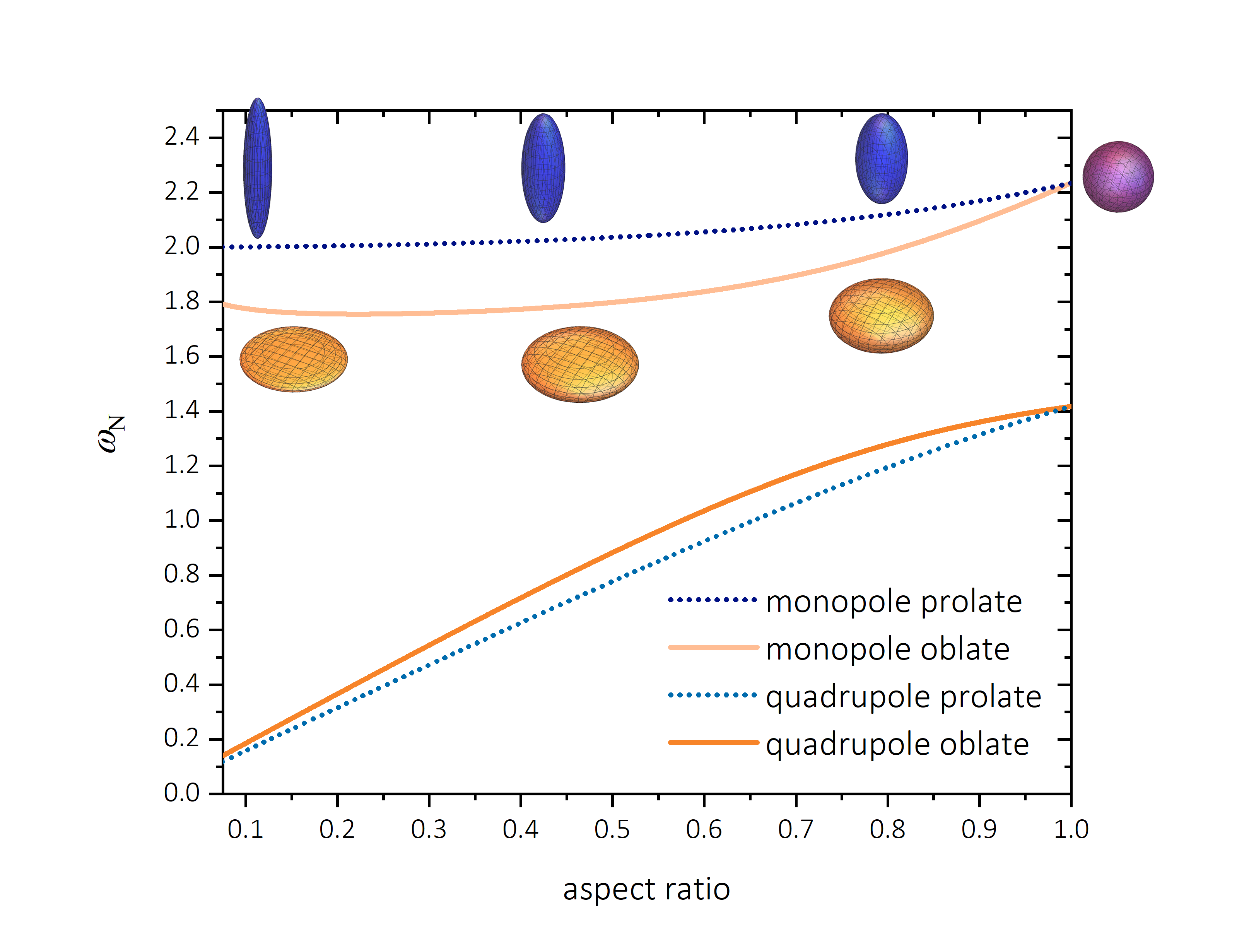}
    \caption{\label{fig:natural_frequencies} Natural frequencies $\omega_b$ and $\omega_q$ for the breathing (monopole)  and quadruple modes, respectively,  obtained by a variational analysis for a Bose-Einstein Condensate in oblate and prolate geometries.  The solutions plotted in this figure were obtained for $a_s=1500 a_0$ and $N=5\times 10^4$.  The higher frequencies correspond to the breathing mode, while the lower frequencies are associated with the quadrupole mode. In the left extreme of the chart, where the aspect ratio is the smallest, the prolate trap is at its maximum confinement in the radial direction, while the oblate trap is at its maximum confinement in the axial direction. As the aspect ratio increases, both traps change their configuration until they reach a spherical geometry, and thus the value of the frequencies in each configuration matches.}
\end{figure}

\begin{table}
\centering
\caption{Definition of the parameters present in Eqs.~\eqref{delta_oblate}~--~\eqref{delta_as} .}
\begin{tabular}{|c | c |} 
 \hline
 \textbf{Modulation} & \textbf{Parameters} \\
 \hline\hline
 \hline
 Prolate & $A_{z}=\left(3\omega_{z0}^{2}+\frac{1}{\sigma_{z0}^{4}}\right),\quad A_{r}=4\omega_{r0}^{2},\quad Q_{r}=-\frac{\alpha \omega_{r0}^{2}}{2},\quad f_{r0}^{\prime}=-\alpha \omega_{r0}^{2} \sigma_{r0}$  \\
$\omega_r$ modulation  &  \\
\hline
 Oblate  & $A_{r}^{\prime}=4\omega_{r0}^{2},\quad A_{z}^{\prime}=\left(3\omega_{z0}^{2}+\frac{1}{\sigma_{z0}^{4}}\right),\quad Q_{z}^{\prime}=-\frac{\alpha\omega_{z0}^{2}}{2},\quad  f_{z0}^{\prime}=-\alpha\omega_{z0}^{2}\sigma_{z0}$   \\ 
$\omega_z$ modulation  &  \\
 \hline
 Prolate and Oblate& $A_{r}^{\prime\prime}=4\omega_{r0}^{2},\quad Q_{r}^{\prime\prime}=-\frac{3}{2}\sqrt{\frac{2}{\pi}}\frac{N\alpha a_{s}}{\sigma_{r0}^{4}\sigma_{z0}},\quad b=\sqrt{\frac{2}{\pi}}\frac{N\alpha a_{s}}{\sigma_{r0}^{3}\sigma_{z0}^{2}},\quad f_{r0}^{\prime\prime}=\sqrt{\frac{2}{\pi}}\frac{N\alpha a_{s}}{\sigma_{r0}^{3}\sigma_{z0}} $ \\
  $a_s$ modulation & $A_{z}^{\prime\prime}=3\omega_{z0}^{2}+\frac{1}{\sigma_{z0}^{4}},\quad Q_{z}^{\prime\prime}=-\sqrt{\frac{2}{\pi}}\frac{N\alpha a_{s}}{\sigma_{r0}^{2}\sigma_{z0}^{3}},\quad f_{z0}^{\prime\prime}=\sqrt{\frac{2}{\pi}}\frac{N\alpha a_{s}}{\sigma_{r0}^{2}\sigma_{z0}^{2}}$\\
  \hline
\end{tabular}
\label{table:1}
\end{table}
\noindent By applying a Fourier Transform (FT) to Eq.~\eqref{matrix_eq}, we get the following matrix equation in the frequency domain, 
\begin{equation}
(\imath \omega_N)^2 \Delta = \mathcal{W}_0 \Delta, \label{matrix_FT} 
\end{equation}
where $\Delta$ is the Fourier transform of vector $\bm{\delta}$. Its diagonalization gives the frequencies of the breathing $\omega_b$ and quadrupolar $\omega_q$ modes, as the eigenvalues of the matrix $\mathcal{W}_0$, i.e., $\omega_N=\{\omega_b,\omega_q \}$. In the dimensionless form they are given by:
\begin{eqnarray}
\omega_{b}=\sqrt{\frac{1}{\sigma_{z0}^4}+4\omega_{r0}^2 + 3\omega_{z0}^2 + \sqrt{\left(\frac{1}{\sigma_{z0}^4}+ 3\omega_{z0}^2-4\omega_{r0}^2 \right)^2 + \frac{16 a_{s0}^2 N^2}{\pi \sigma_{r0}^6 \sigma_{z0}^4} }} \\
\omega_{q}=\sqrt{\frac{1}{\sigma_{z0}^4}+4\omega_{r0}^2 + 3\omega_{z0}^2 - \sqrt{\left(\frac{1}{\sigma_{z0}^4}+ 3\omega_{z0}^2-4\omega_{r0}^2 \right)^2 + \frac{16 a_{s0}^2 N^2}{\pi \sigma_{r0}^6 \sigma_{z0}^4} }} 
\end{eqnarray}
 Figure \ref{fig:natural_frequencies} shows the solution to Eq.~\eqref{matrix_FT} by considering $a_{s0}=1500 a_0$ and $N=5\times 10^4$. As discussed in Section \ref{sec:numericalsimulation}, we consider these values because the simulations performed in this work are based on experimental data of an ultracold fermionic $^6\mathrm{Li}$ gas in the BEC limit \cite{Tey-prl110, Bartenstein-prl92, Toniolo-pra97}. The frequencies of the collective excitations $\omega_N$ are plotted as a function of the aspect ratio for the prolate and oblate trap geometries. As expected, when the aspect ratio is one, the frequencies of each collective excitation are equal for both geometries, which means that our results are consistent in any trap geometry that can be produced by varying the ratios of a cylindrical harmonic trap.  The definition of this frequency hydrodynamic modes are only valid whithin a variation of $N$, $\omega_{r0}$, $\omega_{z0}$, and $a_{s0}$ that keeps valid the use of the GPE equation. 

\begin{figure}[h]
    \includegraphics[width=\textwidth]{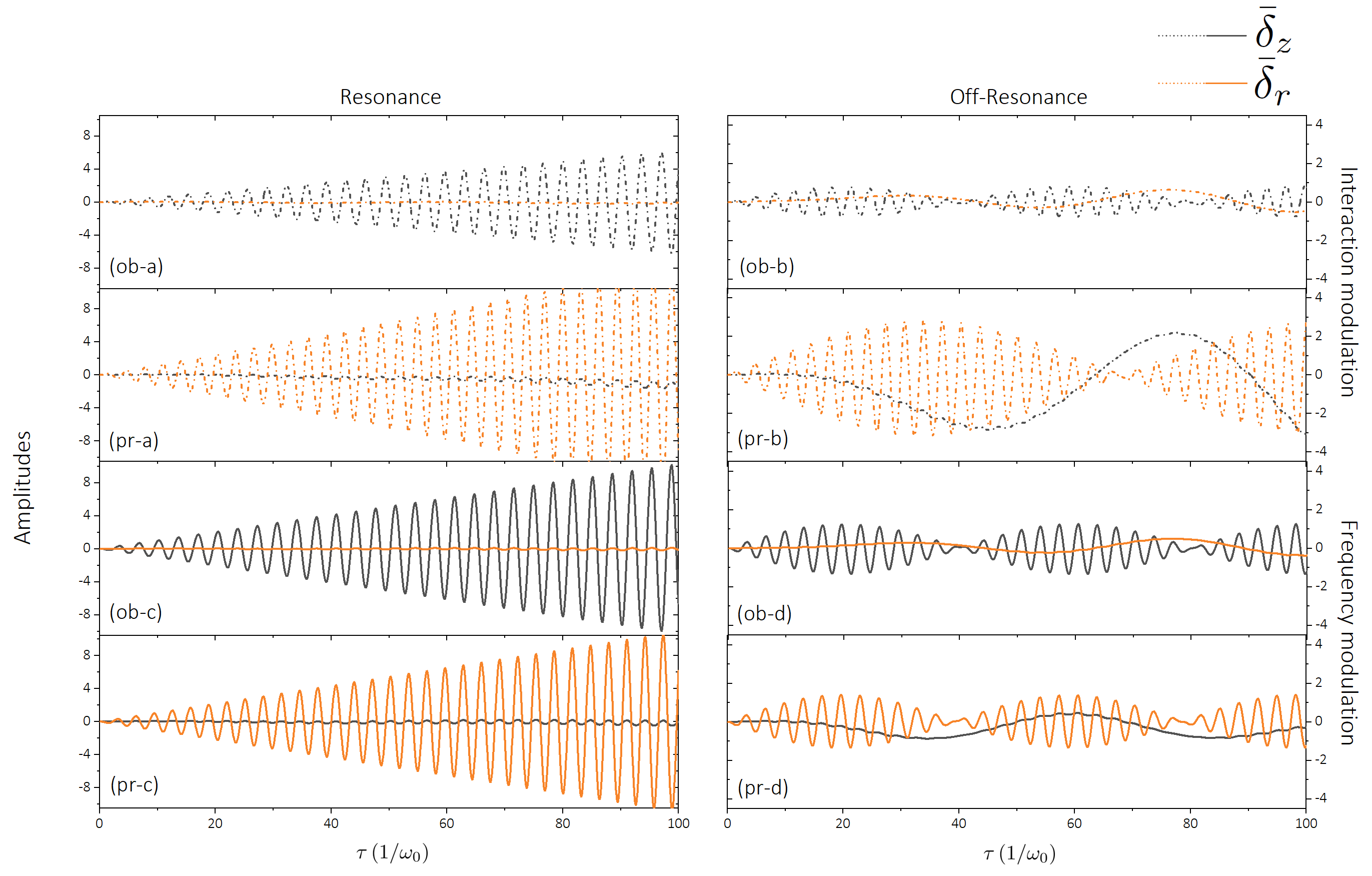} 
    \caption{\label{fig:resonances_traps} Solution to Eqs.~\eqref{delta_prolate}-\eqref{delta_as} in the cases when $\Omega$ takes the value of the breathing mode frequency (obtained with the variational analysis), and a value out of resonance by a small shift in the breathing mode frequency. Time $\tau$ is given in units of $\frac{1}{\omega_0}$, where $\omega_0=\omega_{r0}$ for the prolate trap and $\omega_0=\omega_{z0}$ for the oblate trap. Thus, the modulation for the prolate trap is introduced with a frequency of $\Omega=2.0 \omega_r$ for the case in resonance and $\Omega=(1+0.1)2.0 \omega_r $ for out of resonance. For the oblate trap, the frequency modulation is $\Omega=1.80\omega_z$ for the case in resonance and $\Omega=(1+0.1)1.8 \omega_z$ for out of resonance. Notation (ob-x) and (pr-x) for the subfigures stand for differentiating the geometry of the external trap. The former denotes the case of an oblate trap, whereas the latter denotes a prolate trap.  Subfigures (ob-a), (ob-b), (pr-a), and (pr-b) show the solution to Eqs.~\eqref{delta_as}, when the parametric modulation is introduced through the scattering length. Subfigures (ob-c), (ob-d), (pr-c), and (pr-d) plot the solution to Eqs.~\eqref{delta_oblate} and \eqref{delta_prolate} when the parametric modulation is introduced in the axial or the radial trap frequency, respectively. 
}
\end{figure}

Figure \ref{fig:resonances_traps} shows the solutions to Eqs.~\eqref{delta_prolate}-\eqref{delta_as} in two cases:  when $\Omega$ is equal to the breathing mode $\omega_b$ (obtained with the variational analysis), making the system to oscillate in resonance, and when $\Omega$ is shifted from the breathing mode resonance by a small amount $\epsilon$, $\Omega= \omega_b \pm \epsilon$, but keeping the system inside the width resonance of the Mathieu equation \eqref{mathieueq}, given by $(\omega_{k0}/\kappa)^n$.  Subplots (ob-a), (pr-a), (ob-c), and (pr-c) shows the response of the system when it is modulated at a frequency equal to the breathing mode resonance, $\Omega=\omega_b$. In this case, the system oscillates with high amplitude in the direction of maximum confinement, whereas the perpendicular axis remains little affected.
Subplots (ob-b), (pr-b), (ob-d), and (pr-d) show that a small variation in the value of the resonance frequency (breathing mode) changes the oscillation of the widths in both directions. Thus, a small shift in the resonance frequency, $\Omega=\omega_b \pm \epsilon$, changes the conditions of the system under which we want to study it. As discussed in the following Sections, this small variation in the frequency modulation is also reflected in the periodicity of the temporal patterns obtained by numerical simulations, and in the time at which the patterns appear.  


Summarizing, the variational analysis has been used for  obtaining the natural frequencies of the atomic gas corresponding to the breathing and quadrupole modes. Also, it allows  to get coupled linear differential equations that describe the temporal dynamics of the system in terms of the variational parameters when the condition $\delta_h \ll \sigma_{h0},\, (h=r,z)$ is satisfied. 
Evidently, nearby resonance, the widths $\sigma_h$ are expected to oscillate with increasing amplitude, so at some point the approximation $\delta_h\ll \sigma_{h0}$ could no longer be valid, however, we have ensured that this condition is maintained on the time scale considered in this work when solving Eqs. \eqref{delta_oblate}-\eqref{delta_as}.



\section{Numerical Simulations}\label{sec:numericalsimulation}

We solve numerically the three-dimensional GPE for a  molecular Bose-Einstein Condensate of $^{6}\mathrm{Li}$ with a periodic modulation in the external potential $V(\mathbf{r},t)$ or in the molecular scattering length $a_M(t)$ which determines the interaction strength between the bosonic molecules of mass $M$. The trap frequencies in each configuration are settled close to experimental values or within the ranges of technical accessibility. At $t=0$, when no modulation is applied, the oblate (prolate) trap frequencies are given by $\omega_{r0} = 2 \pi \times 10 ~ \mathrm{Hz} $ and  $\omega_{z0} = 2 \pi \times 150 ~ \mathrm{Hz} $ ($\omega_{r0} = 2 \pi \times 150 ~ \mathrm{Hz} $ and $\omega_{z0} = 2 \pi \times 15 ~ \mathrm{Hz} $). The effective scattering length between the bosonic molecules is given by  $a_{M0}=0.6 \times a_{s0}$ \cite{Petrov-prl93}, where $a_{s0}=1500 a_0$ is the scattering length between the fermionic isotopes of $^{6}\mathrm{Li}$. The number of pairs is fixed to $N = 5 \times 10^4 $.  For these experimental parameters, the dimensionless interaction strength $\eta$ is given by $\eta=1.38$ for the prolate trap, and $\eta=2.32$ for the oblate trap.  The excitation frequency $\Omega$ is $\Omega=2\omega_r$ for the prolate trap and $\Omega=1.80\omega_z$ for the oblate trap, as it was found by the variational method.  The dimensionless variables introduced for solving the GPE  are defined in the same way as in \eqref{eq:holength}, where the harmonic oscillator length $l_0$ is defined through the most confining trap frequency $\omega_0$, equal to $\omega_z$ ($\omega_r$) in the oblate (prolate) geometry. 

All our simulations  were performed using Google's Colab Pro cloud GPU in CUDA language. Eq.~\eqref{GPE} was solved with the Time Split Fourier Method \cite{Chiofalo-pre62, Bao-siam25, Bao-jcompphys187}.  Our system presents cylindrical geometry, hence, in the case of the simulations with the prolate trap, a $256 \times 256 \times 1024 $ grid was used, with an equivalent resolution of $h=0.5 \times 10^{-6} ~  \mathrm{m}$, where $dr=0.5\cdot h$ and $dz=h$, while for the oblate trap, we had a $512\times512\times256$ grid where $dr=h$ and $dz=0.5\cdot h$. This guaranteed a good spatial resolution of the problem once it was normalized with the length of the oscillator corresponding to each of the  considered geometries. On the other hand, the temporal resolution was defined in terms of the excitation frequency $\Omega$, where the time is given  in terms of the number of cycles $N_{\Omega}$, and the total excitation time is given by $T_{\mathrm{exc }} = 2\cdot \pi \cdot N_{\Omega} / \Omega$. The time step is $dt = T_{\mathrm{exc}}/2^{13}$,  and we consider excitation times around 250 ms and 200 ms for the prolate and oblate traps, respectively, both with time steps in the order of $dt =0.4 ~ \mu \mathrm{s}$. These total excitation times are chosen to be long enough for the patterns of interest to form, but still short enough so that the condition $\delta_h\ll \sigma_{h0}$ remains valid.

\section{Parametric Excitation of a BEC: Exact Solutions of GPE}\label{sec:solutions}
 Once we obtain the evolution of the wavefunction of the parametrically driven condensate, we can study the formation of the density patterns. In this section, we discuss our numerical results obtained by solving \eqref{GPE} in four different configurations: i) prolate trap parametrically modulated through the radial trap frequency, ii) prolate trap parametrically modulated through the scattering length, iii) oblate trap parametrically modulated through the axial trap frequency, and iv) oblate trap parametrically modulated through the scattering length. To understand the patterns arising, we decompose the total energy to study the contributions of the kinetic, potential, and interaction terms. In addition, these energy terms are separated into the radial and axial components. The behavior of the energy terms is matched with the patterns observed over the 2D integrated density of the condensate. Also, we  compute the density current to observe the changes the system presents when parametric modulation is introduced. Finally, the temporal periodicity of these patterns and their lifetime are obtained through a fidelity function. 
 
\subsection{Energy Distributions}\label{subec:energies}

\begin{figure}[ht]
\includegraphics[width=\textwidth]{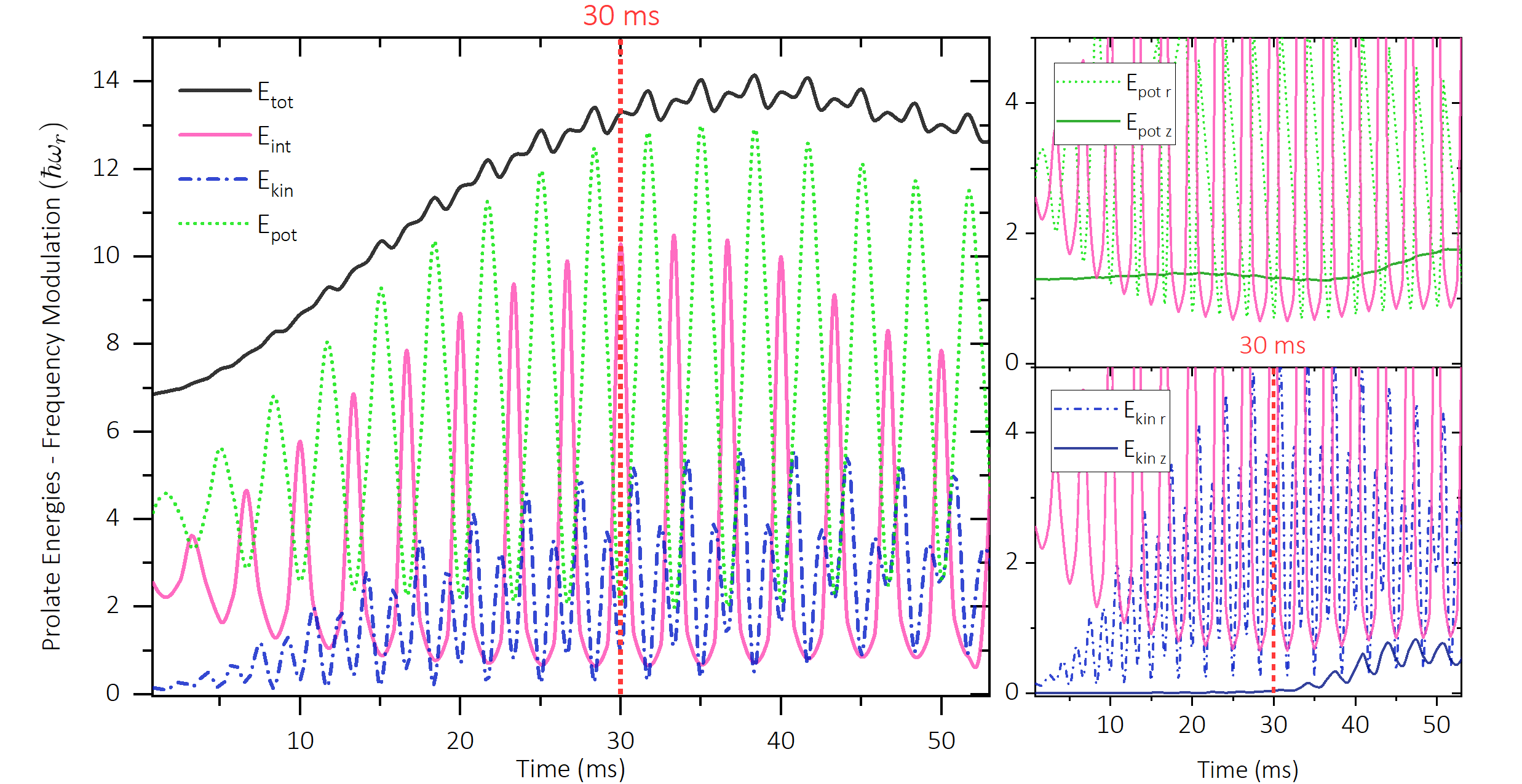}
\caption{\label{fig:Energies_Prolate_FreqMod} Evolution of the system energies for a prolate trap  in which the radial trap frequency is modulated by $\Omega=2.0\,\omega_r$. The plotted energies are given in units of $\hbar \omega_r$.  As the driven system evolves, the total energy  increases, where the principal contributions are given by the $\mathrm{E}_{\mathrm{pot}}$ and $\mathrm{E}_{\mathrm{int}}$. The separation in the axial and radial terms shows that $\mathrm{E}_{\mathrm{kin\,z}}$ remains almost zero at the beginning of the modulation, but after 30 ms, it begins to increase, and $\mathrm{E}_{\mathrm{tot}}$ decreases stabilizing the system,  as indicated by the vertical dashed line. Thus, the redistribution of the energy between axial and radial components is a mechanism to stabilize the system. 
}
\end{figure}

\begin{figure}[ht]
\includegraphics[width=\textwidth]{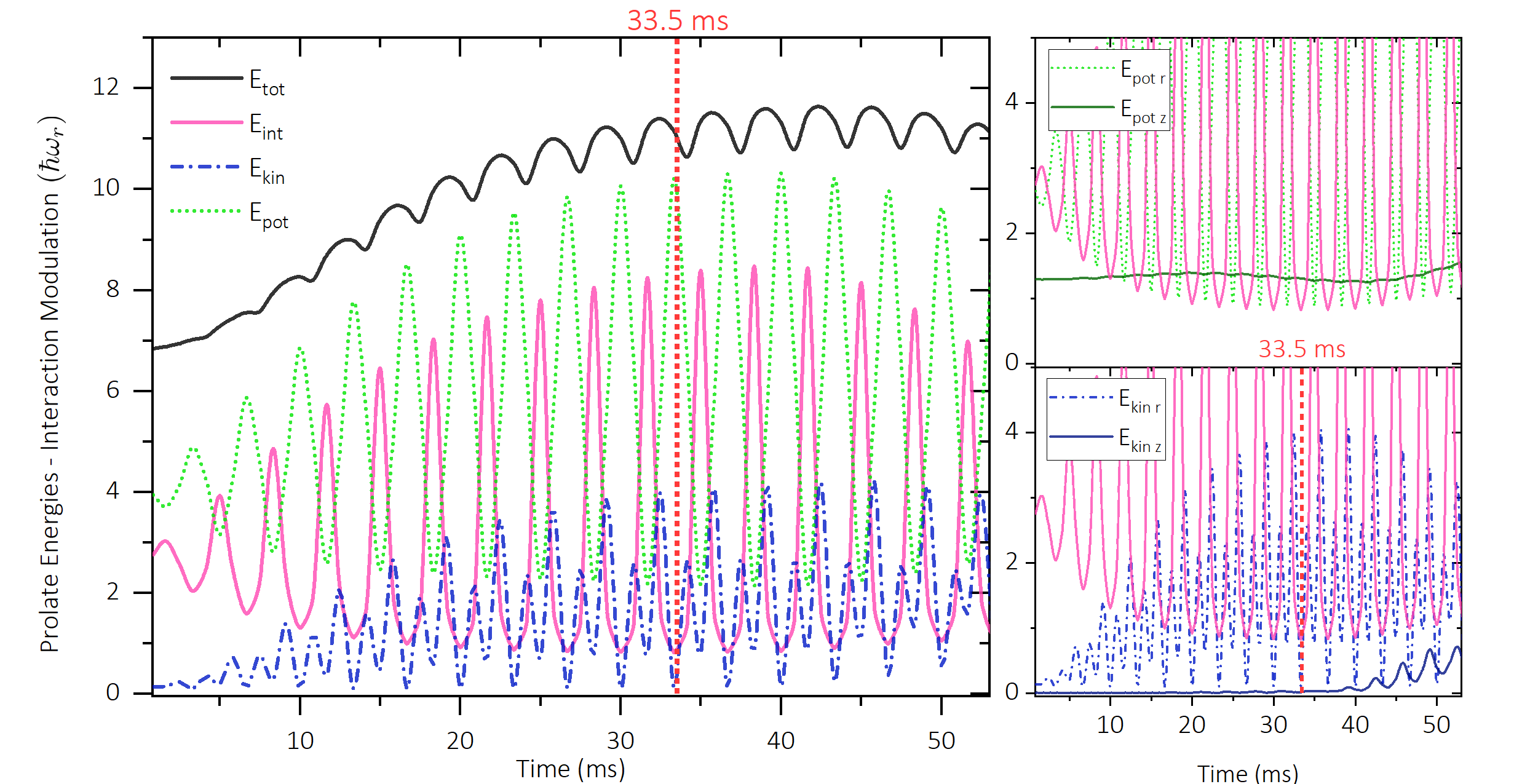}
\caption{\label{fig:Energies_Prolate_IntMod} Evolution of the system energies confined to a prolate trap with a modulation of the scattering length with frequency $\Omega=2.0\,\omega_r$. The plotted energies are given in units of $\hbar \omega_r$. Due to the parametric modulation, an increase in the total energy is observed. In this case, the principal contribution is given by $\mathrm{E}_{\mathrm{pot}}$. In contrast with the parametric modulation through the radial frequency, $\mathrm{E}_{\mathrm{kin}}$  and $\mathrm{E}_{\mathrm{int}}$ take similar values. The separation in the axial and radial terms shows that $\mathrm{E}_{\mathrm{kin\,z}}$ increases after 40 ms, while $\mathrm{E}_{\mathrm{tot}}$ stabilizes and stops increasing  as indicated by the dashed vertical line.  
}
\end{figure}

\begin{figure}[ht]
\includegraphics[width=\textwidth]{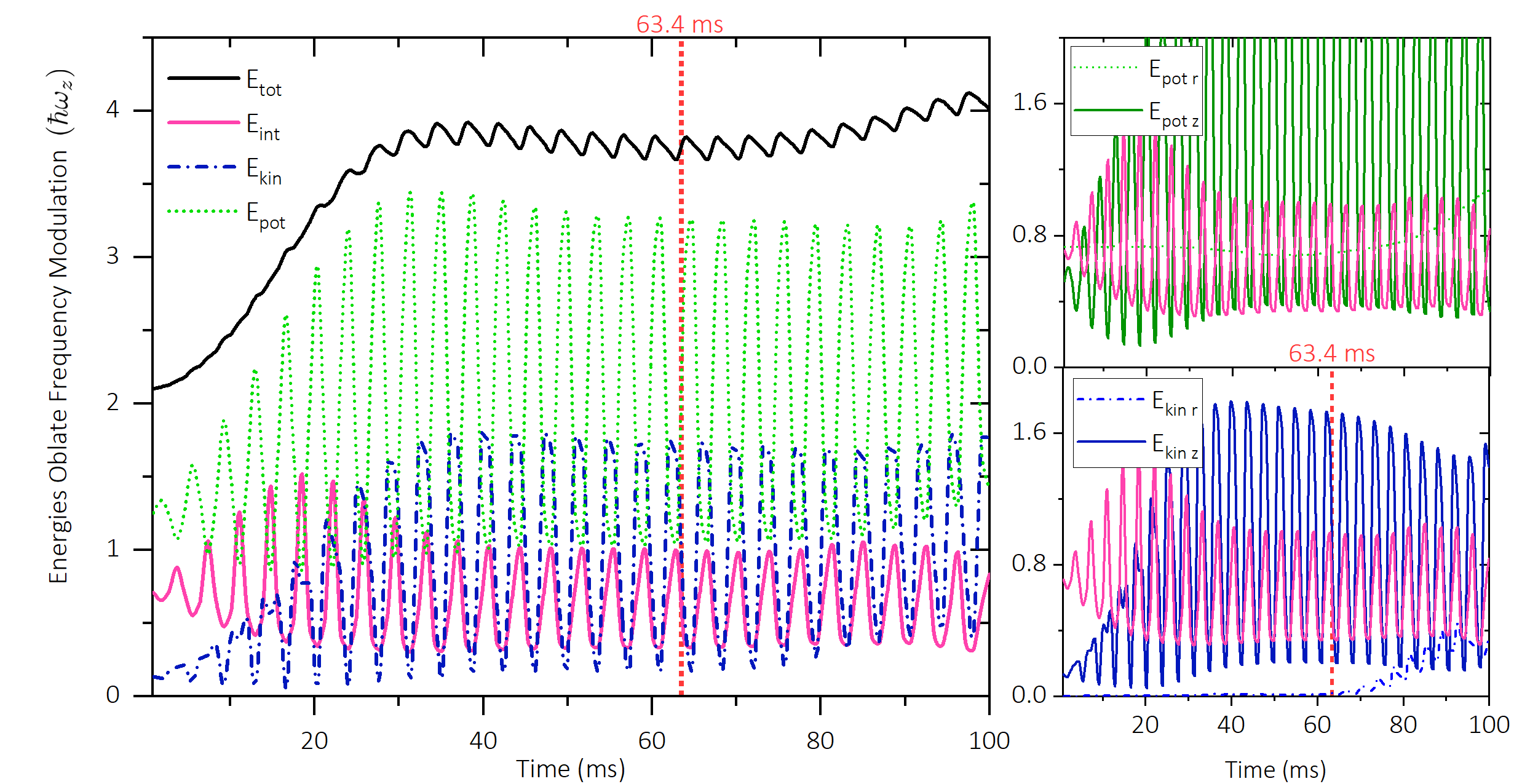}
\caption{\label{fig:Energies_Oblate_FreqMod} Evolution of the system energies for an oblate trap  in which the axial trap frequency is modulated with $\Omega=1.80\,\omega_z$. The plotted energies are given in units of $\hbar \omega_z$. The parametric modulation increases the total energy of the system. The principal contribution is given by $\mathrm{E}_{\mathrm{pot}}$. The separation in the axial and radial terms shows that $\mathrm{E}_{\mathrm{kin\,r}}$ remains almost zero at the beginning of the modulation, but after 60 ms, it begins to increase as indicated by the dashed vertical line. In contrast with the prolate geometry with radial modulation, the total energy does not decrease after the energy's redistribution but continues to increase.
}
\end{figure}

\begin{figure}[ht]
\includegraphics[width=\textwidth]{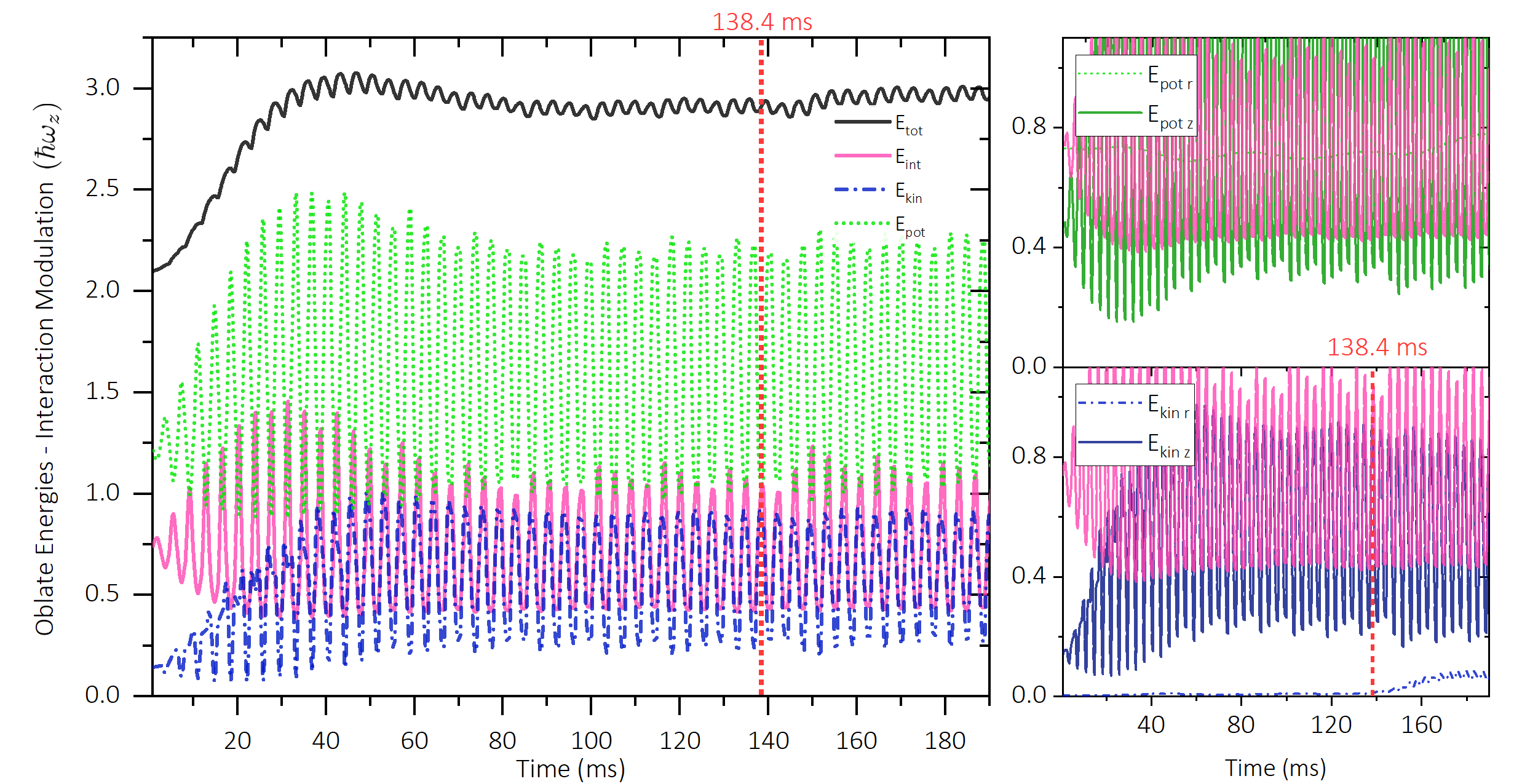}
\caption{\label{fig:Energies_Oblate_IntMod} Evolution of the system energies for an oblate trap in which the axial trap frequency is modulated by $\Omega=1.80\,\omega_z$. The plotted energies are given in units of $\hbar \omega_z$. Due to the parametric modulation, an increase in the total energy is observed. In this case, the principal contribution is given by $\mathrm{E}_{\mathrm{pot}}$. In contrast with the parametric modulation through the radial frequency, $\mathrm{E}_{\mathrm{kin}}$  and $\mathrm{E}_{\mathrm{int}}$ take similar values. The separation in the axial and radial terms shows that $\mathrm{E}_{\mathrm{kin\,r}}$ increases after 140 ms, and $\mathrm{E}_{\mathrm{tot}}$ does not decrease  as indicated by the dashed vertical line.
}
\end{figure}

The total energy of the system is given by 
\begin{equation}
\mathcal{E}(t) = \int d^3 r \,\left\{\right.\underbrace{\frac{\hbar}{2 M}\left| \nabla \Psi(\mathbf{r},t)\right|^2}_{\mathrm{E}_{\mathrm{kin}}} + \underbrace{V_{\mathrm{ext}}(\mathbf{r},t)\left|\Psi(\mathbf{r},t) \right|^2}_{\mathrm{E}_{\mathrm{pot}}} + \underbrace{\frac{1}{2}g(t) \left|\Psi(\mathbf{r},t) \right|^4}_{\mathrm{E}_{\mathrm{int}}} \left.\right\}, \label{total_energy}
\end{equation}
where the time-dependent functions $V_{\mathrm{ext}}(\mathbf{r},t)$ and $g(t)$ are defined in \eqref{eq:potential_mod} and \eqref{eq:gt}, and represent the parametric modulation in one of the trap frequencies (radial frequency $\omega_r(t)$ for the prolate trap or axial frequency $\omega_z(t)$ for the oblate trap) or in the interaction. The different terms in \eqref{total_energy} should be affected differently by the driving. Thus, we begin by comparing the relative values between the kinetic $\mathrm{E}_{\mathrm{kin}}$, potential $\mathrm{E}_{\mathrm{pot}}$, and interacting $\mathrm{E}_{\mathrm{int}}$ energies. In Figures \ref{fig:Energies_Prolate_FreqMod} and \ref{fig:Energies_Prolate_IntMod}, the decomposition of the total energy for the prolate geometry is presented when  the parametric driving is applied through the radial frequency and the scattering length, respectively.  Equivalently, Figures \ref{fig:Energies_Oblate_FreqMod} and \ref{fig:Energies_Oblate_IntMod} show the decomposition of the total energy for an oblate trap with a parametric modulation in the axial trap frequency and the scattering length, respectively. In all these four cases, the higher contribution to the total energy is given by the external potential term, which is defined by the value of the trap frequencies confining the quantum gas, and thus, at $t=0$, represents the range of energy the system acquires because of the confinement. Therefore, the more confining the trap, the higher the potential energy contribution.

Besides, in all cases, when $t =0$, i.e. when no parametric modulation has been introduced, the system satisfies the Thomas-Fermi approximation where the interaction energy is greater than the kinetic energy. Nevertheless, this regime is lost once the modulation begins and the kinetic and interacting energies become comparable. 

Another similitude between the four configurations (geometries and parametric modulations) is found in the axial and radial decomposition of the energy terms. The potential and kinetic energy of the most confined direction, i.e., the radial direction for the prolate and the axial direction for the oblate traps, have higher values in comparison with their perpendicular directions. As the confinement of a system increases, the internal energy also increases. Thus, at the beginning of the parametric modulation, the kinetic and potential energies of the most confined directions contribute to the oscillations of the total energy. In contrast, these energy terms in the less confined directions, i.e., axial direction for the prolate trap and radial direction for the oblate trap, remain nearly constant. In the case of $\mathrm{E}_{\mathrm{kin}}$, its value is almost zero for the elongated axis. Since the parametric modulation continuously introduces energy to the system, at some point, the system is saturated, and a mechanism to stabilize it is to  transfer energy to the less  confining directions. In this way, we observe the increase of the kinetic and potential energies in the axial direction for the prolate trap (Figures \ref{fig:Energies_Prolate_FreqMod} and \ref{fig:Energies_Prolate_IntMod}). The same energy terms increase in the radial direction for the oblate trap (Figures \ref{fig:Energies_Oblate_FreqMod} and \ref{fig:Energies_Oblate_IntMod}). Notice that in the case of the prolate trap, this energy redistribution is also observed in a decrease in the total energy. Nevertheless, this decrease in $\mathrm{E}_{\mathrm{tot}}$ is not presented for the oblate trap, but the total energy is stabilized. 

The time at which this redistribution occurs depends on the parametric modulation and the trap geometry. For the same geometry (prolate or oblate), the increase of the kinetic component in the less confined direction appears sooner when the parametric modulation is introduced through one of the trap frequencies. In the energy decomposition of the prolate trap, we observe that for the frequency modulation (right subfigures in Figure \ref{fig:Energies_Prolate_FreqMod}), the kinetic energy in the axial coordinate increases after 30 ms, while for the scattering modulation (right subfigures in Figure \ref{fig:Energies_Prolate_IntMod}), this increase is observed after 40 ms. For the oblate trap, the kinetic energy in the radial trap increases after 70 ms for the axial modulation (Figure \ref{fig:Energies_Oblate_FreqMod}) and after 140 ms for the scattering length modulation (Figure \ref{fig:Energies_Oblate_IntMod}).

Another difference between both parametric modulations is the total energy  that is transferred into the system. The system gets higher energy values for the modulation through one trap frequency. This is explained by the fact that the excitations are introduced directly in the most confined direction. In this sense,  the parametric modulation through one trap frequency  is anistropic. In contrast, the modulation of the scattering length has an isotropic nature since it creates excitations in all directions. Thus, within this excitation mechanism, systems with an anisotropic modulation acquire high energies, which are soon redistributed to achieve a stable regime.

\begin{figure}[ht]
\includegraphics[width=\textwidth]{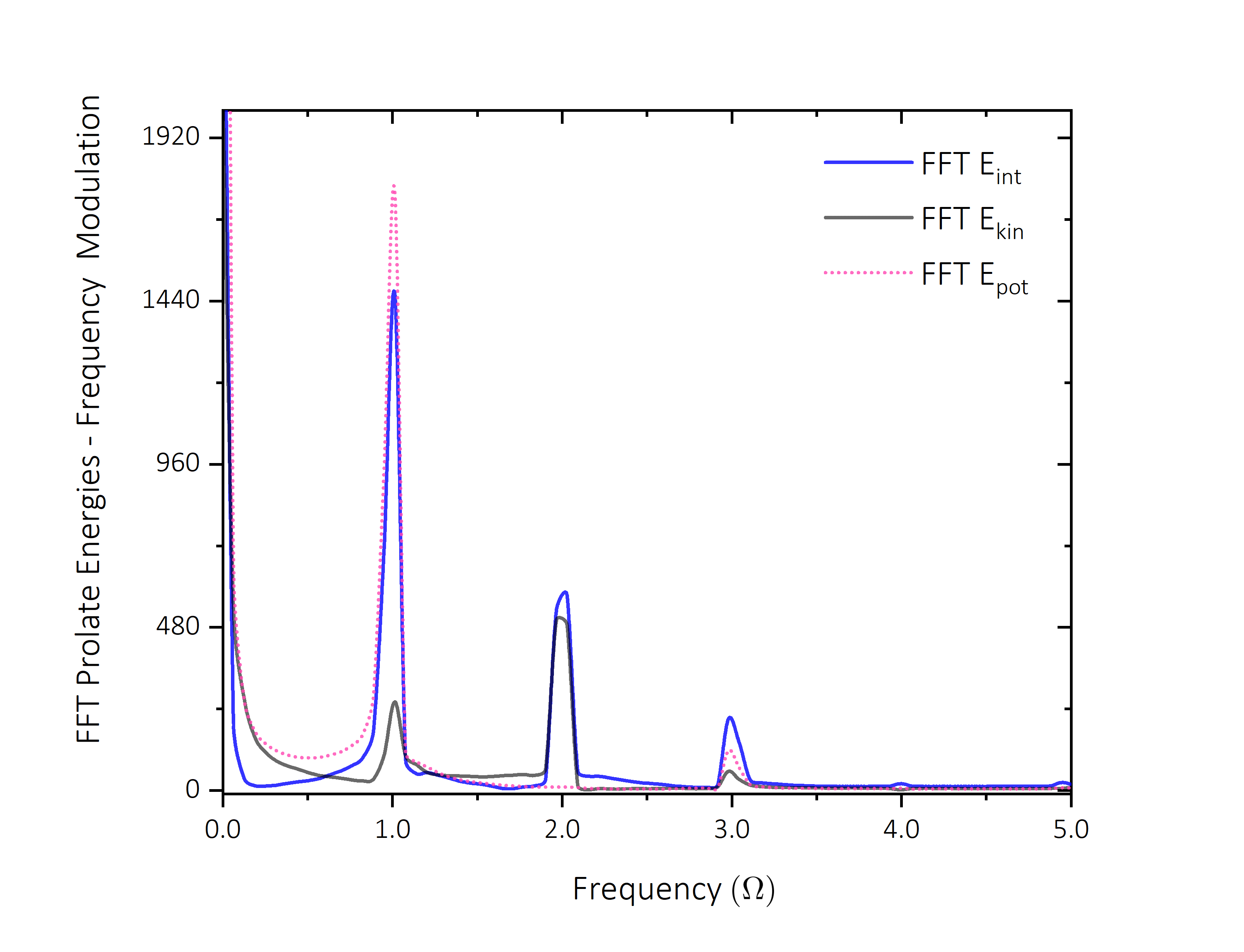}
\caption{\label{fig:FFT_Energies_Prolate_FreqMod}  FFT of the energy terms in \eqref{total_energy} for a prolate trap parametrically modulated through the radial trap frequency. The $x$ axis is in units of the frequency of modulation $\Omega= 2.0 \,\omega_r$. $\mathrm{E}_{\mathrm{int}}$ and $\mathrm{E}_{\mathrm{pot}}$ oscillates principally with a frequency equally to $\Omega$, while $\mathrm{E}_{\mathrm{kin}}$ oscillates mainly with $2\Omega$. 
}
\end{figure}

\begin{figure}[ht]
\includegraphics[width=\textwidth]{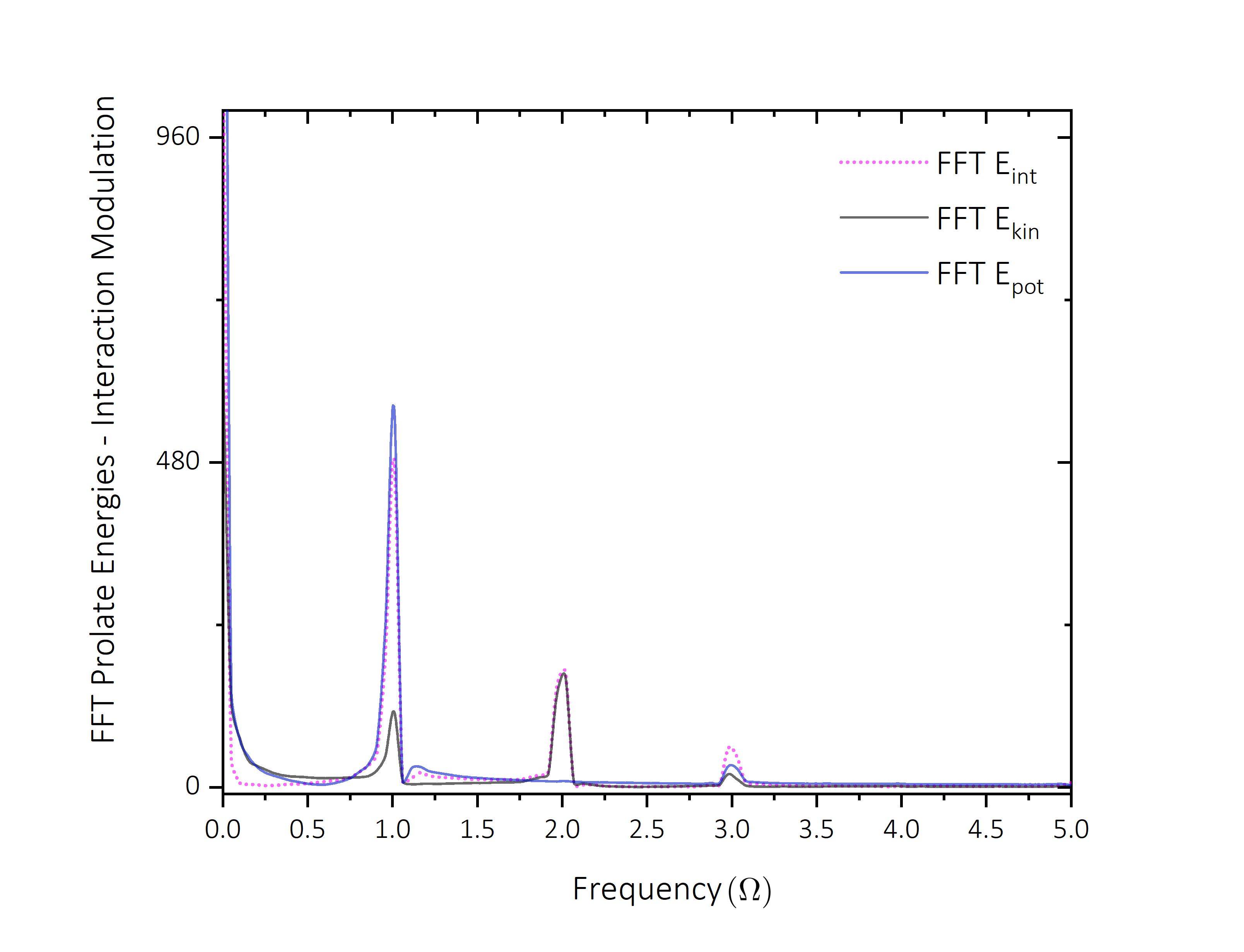}
\caption{\label{fig:FFT_Energies_Prolate_IntMod}  FFT of the energy terms in \eqref{total_energy} for a prolate trap parametrically modulated through the scattering length. The $x$ axis is in units of the frequency of modulation $\Omega= 2.0 \,\omega_r$. In this case, all the energy terms oscillates principally with a frequency equally to $\Omega$.
}
\end{figure}

\begin{figure}[ht]
\includegraphics[width=\textwidth]{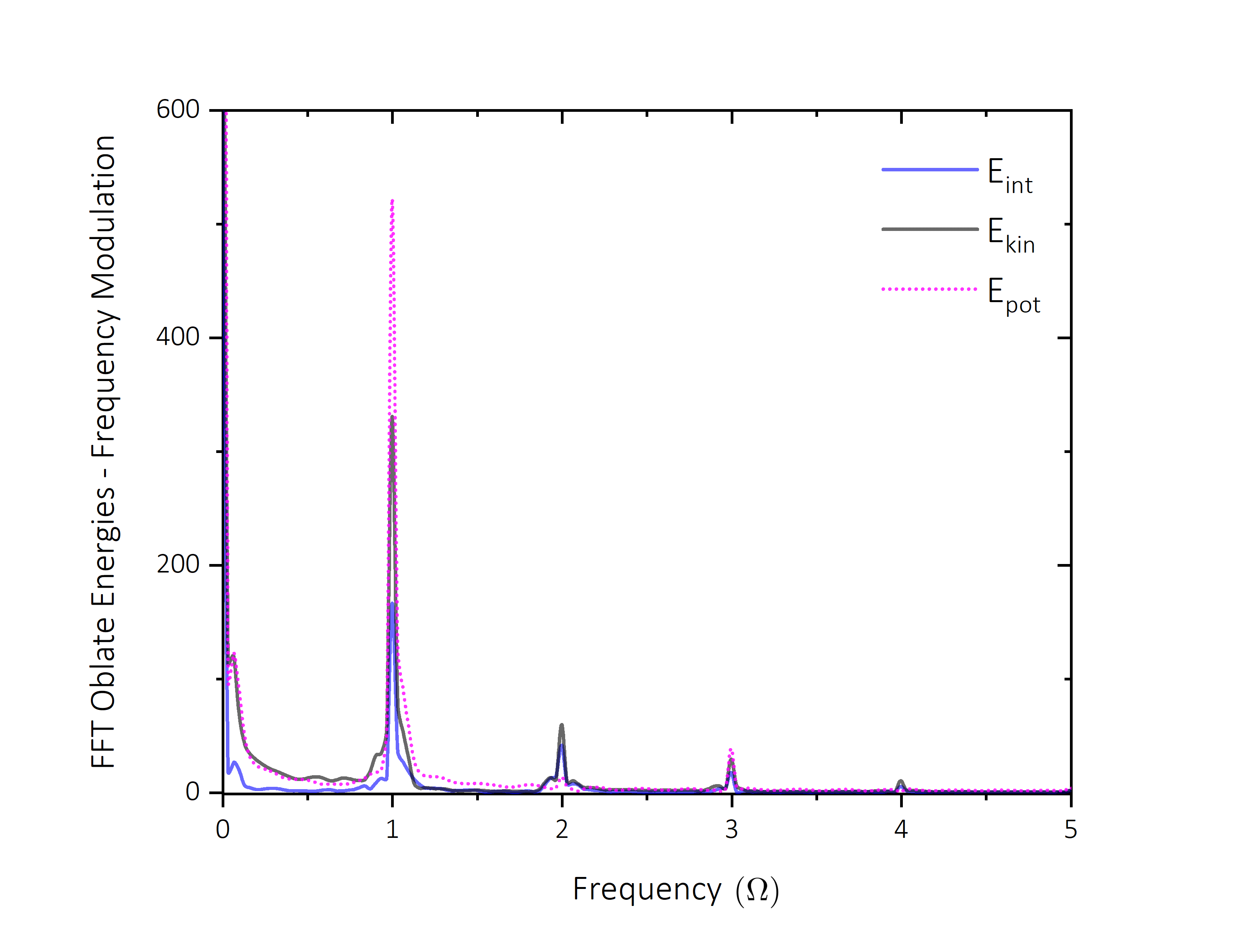}
\caption{\label{fig:GFFT_Oblate_FreqMod} Fast Fourier Transform of the energy terms in \eqref{total_energy} for an oblate trap parametrically modulated through the axial trap frequency. The $x$ axis is in units of the frequency of modulation $\Omega= 1.80 \,\omega_z$. $\mathrm{E}_{\mathrm{int}}$, $\mathrm{E}_{\mathrm{pot}}$ and $\mathrm{E}_{\mathrm{kin}}$ oscillates mainly with a frequency equally to $\Omega$. 
}
\end{figure}

\begin{figure}[ht]
\includegraphics[width=\textwidth]{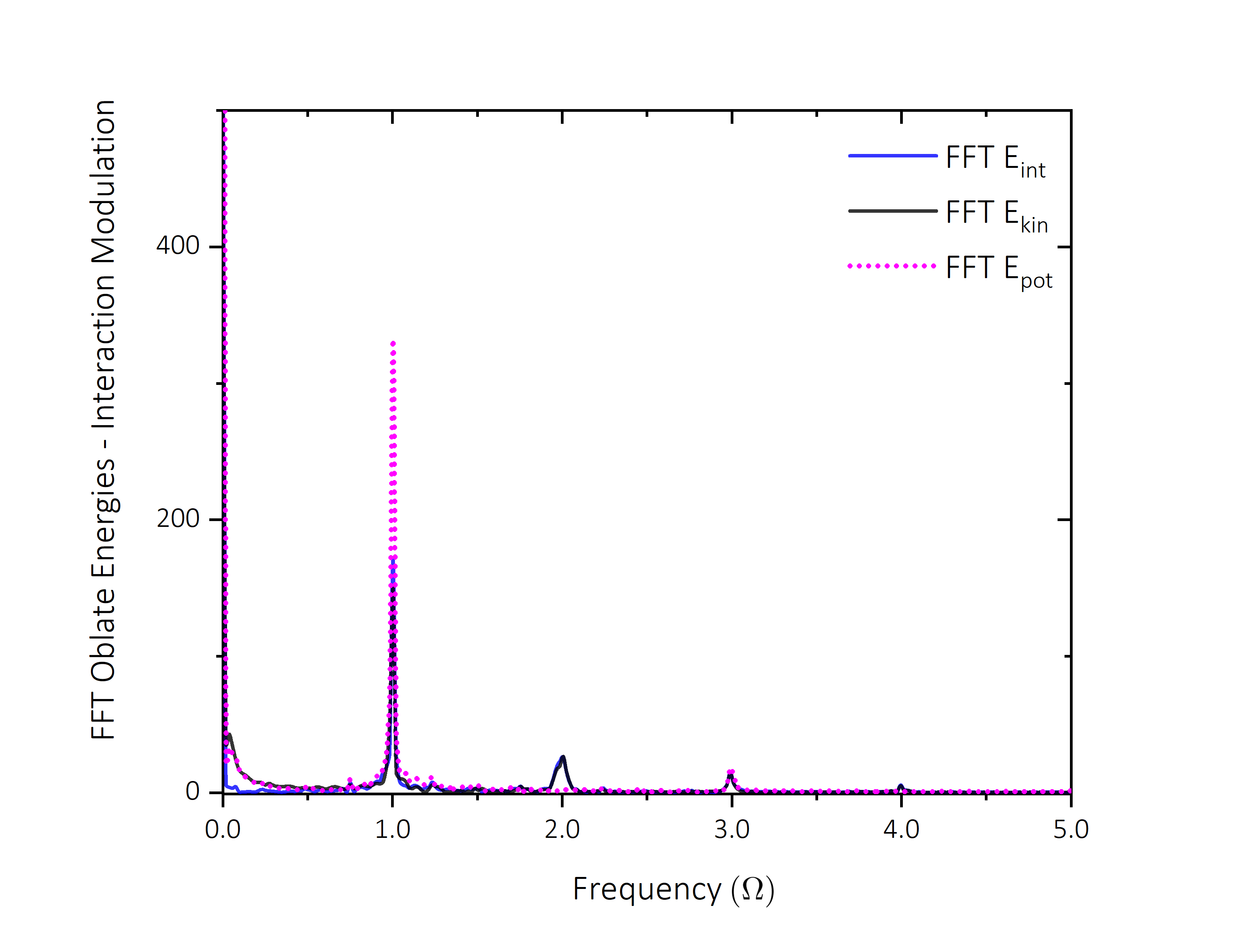}
\caption{\label{fig:GFFT_Oblate_IntMod} Fast Fourier Transform of the energy terms in \eqref{total_energy} for an oblate trap parametrically modulated through the scattering length. The $x$ axis is in units of the frequency of modulation $\Omega= 1.80 \,\omega_z$. $\mathrm{E}_{\mathrm{int}}$, $\mathrm{E}_{\mathrm{pot}}$ and $\mathrm{E}_{\mathrm{kin}}$ oscillates mainly with a frequency equally to $\Omega$. }
\end{figure}

Now, we discuss the periodicity of the energy oscillations. Figures \ref{fig:FFT_Energies_Prolate_FreqMod}  and \ref{fig:FFT_Energies_Prolate_IntMod} show the Fast Fourier Transform (FFT) of the energy decomposition for the prolate trap with a parametric modulation through the radial frequency and the scattering length, respectively. Similarly, the FFT of the energies of the oblate trap with a parametric modulation via the axial frequency and the scattering length are plotted in Figures \ref{fig:GFFT_Oblate_FreqMod} and \ref{fig:GFFT_Oblate_IntMod}, respectively. In all the cases, the plotted frequencies are in units of the excitation frequency $\Omega$, which is $\Omega=2\omega_r$ for the prolate trap and $\Omega=1.80\omega_z$ for the oblate trap. 

Some important differences in the FFT of the energies of the prolate trap (Figures \ref{fig:FFT_Energies_Prolate_FreqMod} and \ref{fig:FFT_Energies_Prolate_IntMod}) are observed. When the parametric modulation is introduced through $\omega_r$, energies $\mathrm{E}_{\mathrm{int}}$ and $\mathrm{E}_{\mathrm{pot}}$ mainly oscillate with a frequency equal to $\Omega$. Nevertheless $\mathrm{E}_{ \mathrm{kin}}$ oscillates principally with $2\Omega$. This behavior is not observed when the parametric drive is at the scattering length, where all the energies oscillate with the same frequency $\Omega$, but some subharmonic contributions appear. 

The FFT of the energies of the oblate trap with a parametric modulation through $\omega_z$ (Figure \ref{fig:GFFT_Oblate_FreqMod}) shows that all the energy terms oscillate mainly with the frequency of modulation $\Omega$, and there are several contributions of subharmonics, in contrast with the case of the prolate trap with the same parametric modulation. For the oblate trap with a modulation in the scattering length (Figure \ref{fig:GFFT_Oblate_IntMod}), the FFT of the energy terms shows that the main contribution to the oscillations is given by $\Omega$. 


\subsection{Density Patterns}

The increasing of the kinetic energy in the less confined directions, discussed in the previous Section \ref{subec:energies},  should be reflected in the local density of the fluid. To understand how it is affected, we plot the 2D integrated density over the most confined direction. This is, we integrate over the z-axis in the oblate trap to plot the xy-plane density. For the prolate trap, we integrate over the y-axis to plot the xz-plane density. We present these integrated densities, since this is the information that can be obtained experimentally through a typically absorption image. 

\begin{figure}[ht]
\includegraphics[width=0.6\textwidth]{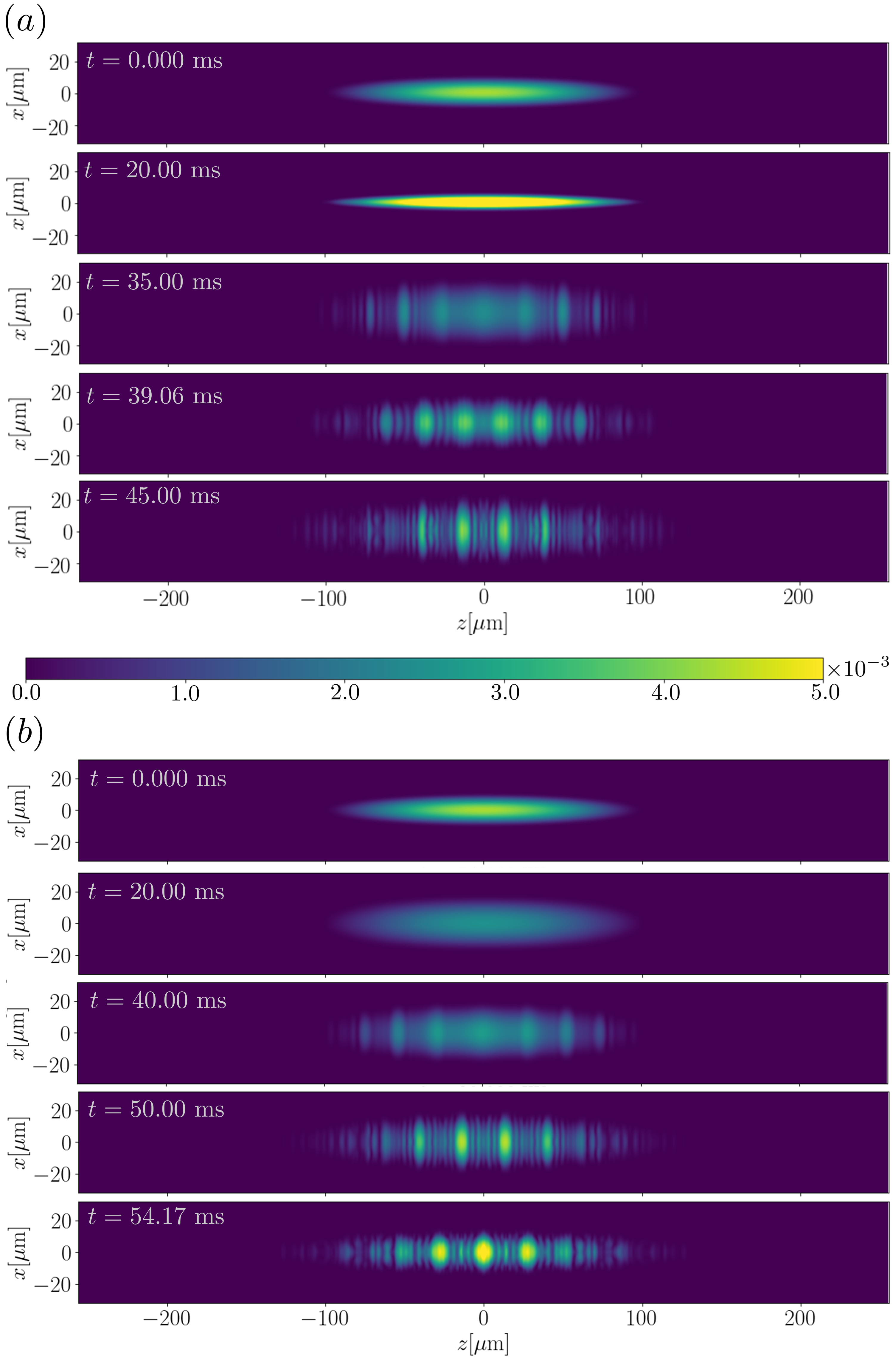}
\caption{\label{fig:LobePatternt0} Time evolution of the 2D integrated density of a prolate trap with a parametric modulation through (a) the radial trap frequency, and (b) the scattering length. The total density is integrated over $y$-plane. This integrated density is comparable to the image density obtained  experimentally through an absorption image. The formation of the density patterns coincides with the time at which the axial kinetic energy $\mathrm{E}_{\mathrm{kin\,z}}$ increases. This increase is observed after 30 ms for the radial trap modulation, and for the scattering length modulation is observed after 40 ms. In both cases, before the time of  this increase, the initial distribution of the density remains unchanged, and the breathing mode oscillation is observed in the contraction and expansion of the atomic cloud. After the axial kinetic energy increases, the density patterns appear and become well-defined when $\mathrm{E}_{\mathrm{kin\, z}}$ takes its maximum values. The formation of the density patterns breaks the initial Thomas-Fermi density distribution and generates fringes with maximum density.}
\end{figure}

Figure \ref{fig:LobePatternt0} shows the formation of fringe patterns over the $xz$ density when a prolate trap is parametrically modulated through the radial trap frequency and the scattering length. As discussed in the previous Section, the kinetic energy in the axial direction begins to increase after 30-35 ms of excitation when the radial frequency of the trap is modulated; or after 40 ms of excitation in the case in which the scattering length is modulated. Before this time, the breathing mode oscillation is observed in both cases due to the parametric modulation with frequency $\Omega=2.0 \,\omega_r$. When kinetic energy in the axial direction increases, a breaking in the prolate density distribution is observed since a faint pattern formation  starts to emerge. These patterns become more  noticeable  after the kinetic energy increases and, as consequence,  the density distribution exhibits a spatially and temporary periodic fringe pattern.

\begin{figure}[ht]
\includegraphics[width=0.55\textwidth]{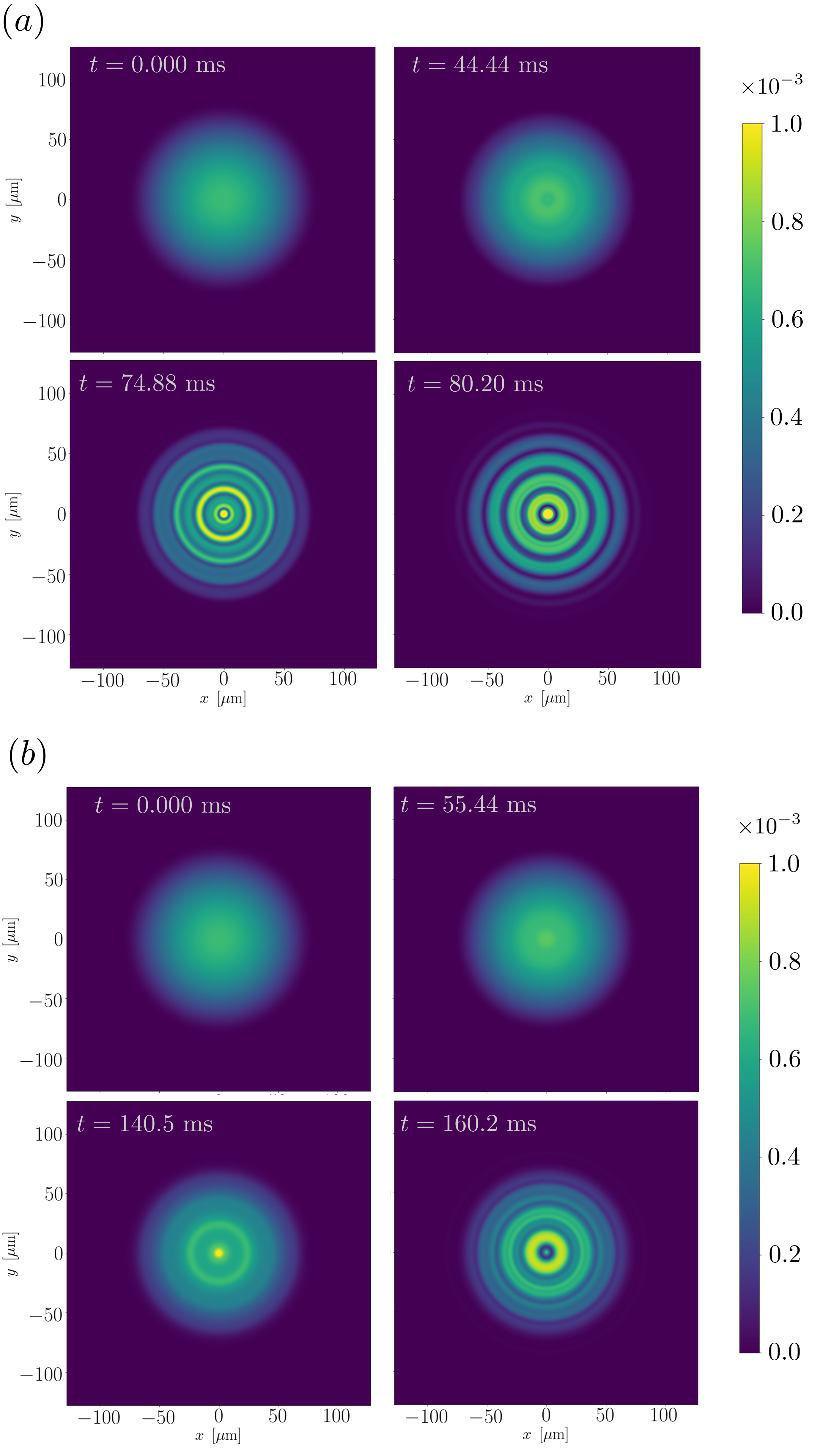}
\caption{\label{fig:OblatePatternsFrequency} Time evolution of the 2D integrated density of an oblate trap with a parametric modulation through (a) the axial trap frequency and (b) scattering length. The total density is integrated over $z$-plane. This integrated density is comparable to the image density obtained experimentally through an absorption image. The formation of the density patterns coincides with the time at which the radial kinetic energy $\mathrm{E}_{\mathrm{kin\,r}}$ increases. For the axial trap modulation, the increment in the kinetic energy is observed after 65 ms, while for the scattering length modulation, the increment is observed after 140 ms. In both cases, faint ring patterns can be observed before the increment, which gets well-defined after the increase of $\mathrm{E}_{\mathrm{kin\, r}}$. Forming the density patterns breaks the initial Thomas-Fermi density distribution and generates rings with maximum density.  
}
\end{figure}

Figure \ref{fig:OblatePatternsFrequency} shows the formation of patterns over the $xy$ density for an oblate trap with a modulation in the axial trap frequency and in the scattering length. A ring pattern formation is observed in these cases due to the trap geometry. Similarly to the prolate case, before the time when there is an increase in the radial kinetic energy, an expansion and contraction of the atomic cloud in the radius of the $xy$ plane is observed due to the breathing mode oscillation induced by the parametric modulation with frequency $\omega=1.80 \, \omega_z$. Also, a faint ring pattern can be observed, which intensifies when the radial kinetic energy increases at $t-t_0\sim 70$ ms for the axial trap modulation and at $t-t_0\sim 140$ ms for the scattering length modulation. Once the ring patterns are completely formed, the simulations show that the characteristic Thomas-Fermi profile, where the maximum density is localized at the center and decreases as we get away from it, is lost. Now, there are different maxima of the density localized within the rings.

Besides, both the prolate and oblate traps simulations show that the density maxima are not static; for different times, they are localized at different positions. As we discussed in the next Subsection, the current density gives information about the density displacement.  

Now, we want to emphasize that the formation of fringe patterns in the prolate trap and ring patterns in the oblate trap match the time of the energy redistribution between radial and axial components. This shows the importance of considering a 3D system for studying these pattern formations since the collective modes in radial and axial directions are correlated.

 To study this phenomenon in a system with fewer dimensions, a potential trap that is  much more elongated in one of the axial or radial directions is needed so that condition \cite{Kamchatnov2004},
 \begin{equation}
\frac{N a_s}{l_0} \ll \left( \frac{R_{TF}}{l_0}\right)^2,
\end{equation}
with $l_0$ defined in \eqref{eq:holength} and $R_{TF}$ as the Thomas-Fermi Radius of the plane of interest, is satisfied. In this dimensional reduction, the redistribution mechanism of the total energy that generates the emergence of the density patterns should be different and interesting to study.

\subsection{Current Density}
\label{CurrentDensity}

As shown in the 2D integrated density plots (Figures \ref{fig:LobePatternt0} and \ref{fig:OblatePatternsFrequency}), the parametric modulation induces a density flow, which breaks the characteristic Thomas-Fermi profile of the condensate. This creates  periodic patterns with maxima density  localized at different positions of the atomic cloud. To understand this process, we compute the condensate current density, a mechanism for redistributing  energy in an ultracold sample with periodic modulation. The current density $\mathbf{J}$ of a system with a wavefunction $\Psi=\sqrt{\rho}\mathrm{e}^{\imath S(\mathbf{r})}$ gives information about the phonon propagation \cite{Capuzzi-pra78}. Recalling its definition, $\mathbf{J}=\rho \mathbf{v}$, where $\rho$ is the density of the system and $\mathbf{v}$ the field of velocities of the superfluid given by $\mathbf{v}=\bm{\nabla}S(\mathbf{r})$, we can rewrite it as $\mathbf{J}=\frac{\rho}{m}\mathbf{p}=\frac{\hbar \rho}{m}\mathbf{k}$, where $\mathbf{k}$ is the local wave vector of the phonon excitations.

The density currents presented in this Section are computed considering only the $xy$ plane that crosses the middle of the atomic cloud in the oblate trap and the $xz$ plane that crosses the middle of the atomic cloud with prolate geometry. 

\begin{figure}[h]
\includegraphics[width=\textwidth]{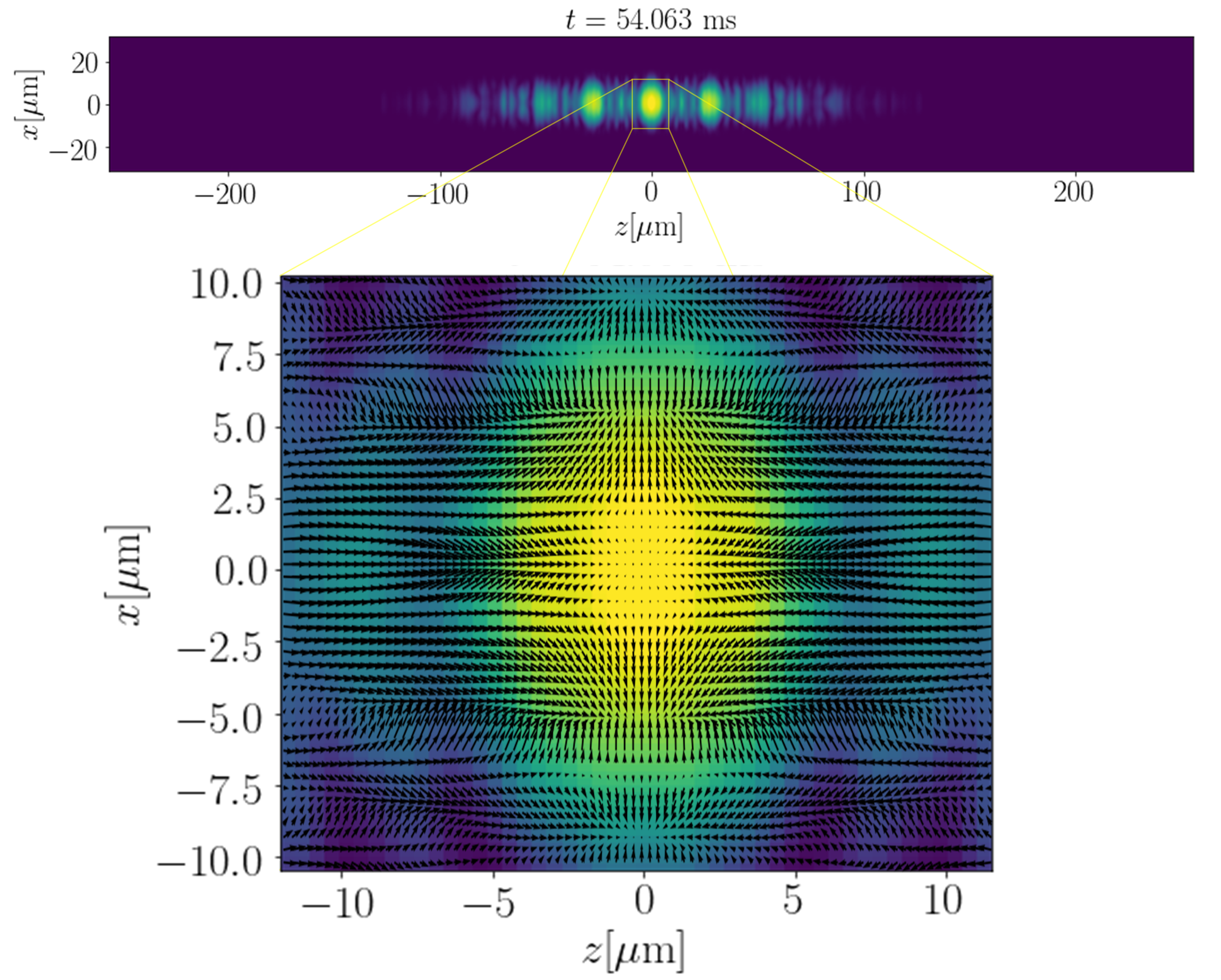}
\caption{\label{fig:CurrentsProlate}Current density of the prolate trap parametrically modulated through the scattering length. The vector field is obtained for the $xz$-plane that crosses through the middle of the atomic cloud. The formation of the fringe patterns, i.e., the localization of maximum densities in fringe patterns through the atomic cloud, is generated due to the interference of the phonons traveling to the extremes of the trap.  }
\end{figure}

Figure \ref{fig:CurrentsProlate} shows the current density behavior in a prolate trap. When the prolate trap is expanded due to the breathing mode collective oscillation, the direction of the vector field indicates that the density is displaced  from the center of the trap to the left and right ends of the z-axis. When the atomic cloud is contracted due to the breathing mode, the direction of the current density points to return to the trap center. Eventually, this mechanism of expansion and contraction of the atomic cloud generating the density to travel away from the trap center and to return to it, generates the fringe patterns when two sections of densities with opposite directions interfere. Interestingly, the geometry of the trap defines the main direction in which the current density travels.

\begin{figure}[h]
\includegraphics[width=0.65\textwidth]{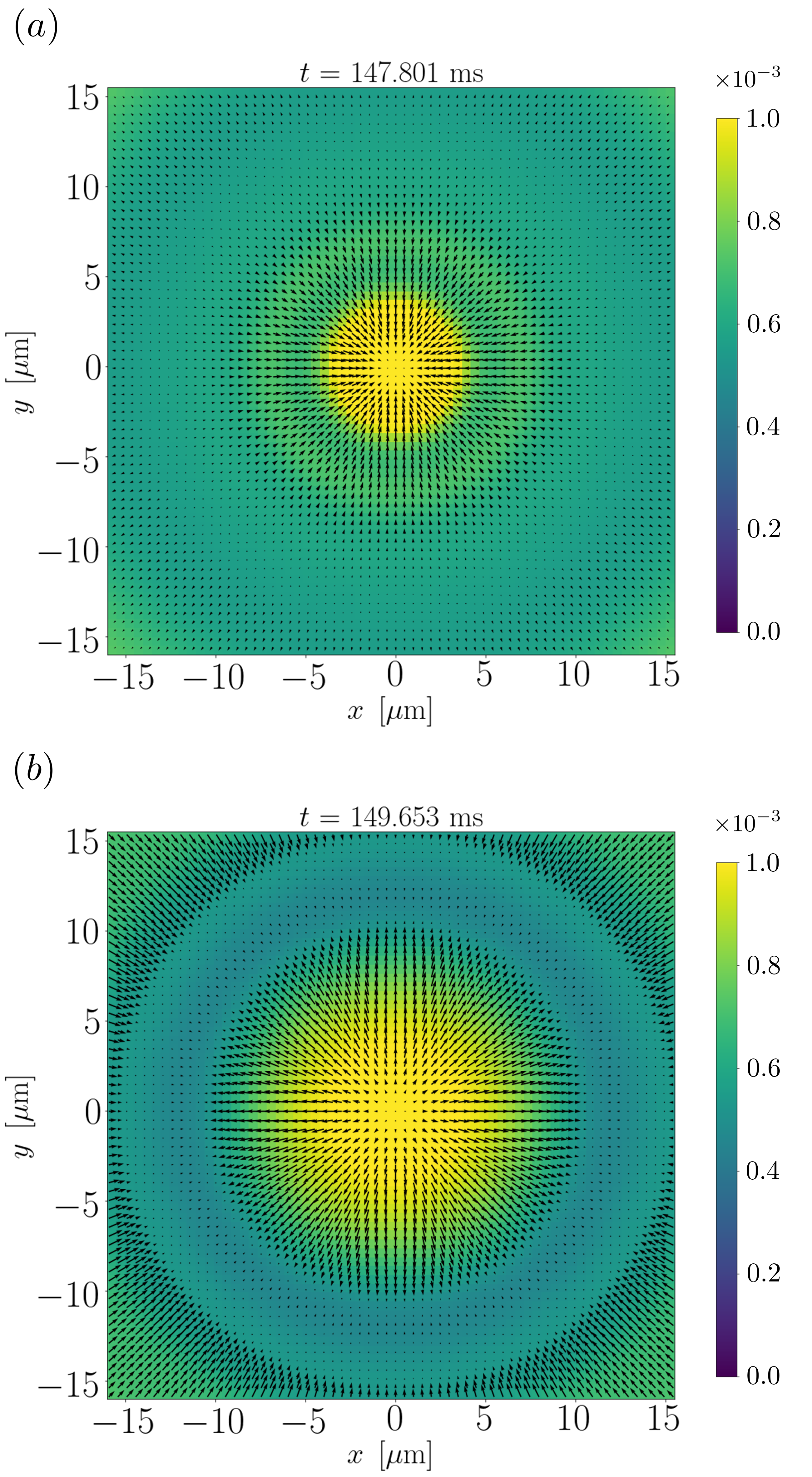}
\caption{\label{fig:Currents1} Current density of the oblate trap parametrically modulated through the scattering length. The vector field is obtained for the $ xy$ plane that crosses through the middle of the atomic cloud. The formation of the ring patterns is generated due to the interference of the phonons traveling in radial direction to the extremes of the trap.  }
\end{figure}

In Figure  \ref{fig:Currents1}, the propagation of the current density is plotted for the oblate trap. Similarly to the prolate case, when the atomic cloud is contracted due to the breathing mode oscillations, the displacement of the density is in a radial direction towards the center of the trap (Fig. \ref{fig:Currents1}(a)). In contrast, when the atomic cloud is expanded, the density is displaced far away from the center traps with radial propagation (Fig. \ref{fig:Currents1}(b)). Again, these density displacements with radial direction originate the ring patterns. 

It is important to note that the local density of angular momenta is null in either of the cases. Therefore, no vortex or other associated topological defect is present in the condensate density. This is to be expected since in our ansatz for the ground state of the wavefunction, \eqref{GaussianAnsatz}, the case with $l = 0$ was considered. However, when the parametric modulation is introduced, the system evolves with propagating excitations that travel in directions that minimize energy but respect the symmetry of $l=0$. In this sense, and due to our trap geometry,  in the oblate trap, the phonons propagate in the radial direction, generating ring patterns. In the prolate trap, the phonons propagate along the $z$-axis generating the fringe patterns. Thus, the external potential imposes the symmetry of the generated patterns. Since our harmonic trap has cylindrical symmetry, the reflection of the excitations preserves the same configuration. Besides, a natural squared symmetry is expected when there is no external potential. This is observed in reference \cite{Fujii2024} where,  by imposing a square symmetry through the ansatz of the wavefunction, they generate stable squared patterns in an infinitely extended BEC without trapping potential and with the scattering length modulated. Since the external potential imposes the symmetry of the system, in our case, the generated patterns (rings in the oblate trap and fringes in the prolate trap) arise naturally and are the reflection of the excitations over the boundaries of the potential. 


\subsection{Fidelity}
The temporal periodicity of the patterns and their lifetimes (the interval of time at which they can be experimentally observed) can be studied by a fidelity function. For simplicity, and similar to the current density discussed in Subsection \ref{CurrentDensity},  we define this function over a plane and not in all the atomic cloud space. Thus, for the prolate trap, we took the $xz$ plane at $y=0$. In the same way, the  $xy$ plane at $z=0$ is taken for the oblate trap. 

The definition of the fidelity function is:
\begin{equation}
F\left[t\right]=\frac{\left| \left\langle \Psi_{ab}^\dagger\left(t+dt\right)\Psi_{ab}\left(t\right)\right\rangle \right|^2}{\left\langle \Psi_{ab}^\dagger\left(t+dt\right)\Psi_{ab}\left(t+dt\right) \right\rangle \left\langle \Psi_{ab}^\dagger\left(t\right)\Psi_{ab}\left(t\right) \right\rangle} \label{FidelityDef}
\end{equation}
where $\Psi_{ab}$ denotes the wavefunction of the plane of interest, i.e., $xz$ plane for the prolate and $xy$ for the oblate trap. Notice that the function is normalized just for the considered plane, therefore, brackets $\langle \rangle$ denote the integration just over the plane. 

To study the time behavior of the generated patterns in both geometries and parametric modulations, we take the wavefunctions at the plane of interest, $\Psi_{xz}$ and $\Psi_{xy}$, evaluated at a time at which the kinetic energy in the less confined direction increases. These times are reported in Figures \ref{fig:Energies_Prolate_FreqMod}, \ref{fig:Energies_Prolate_IntMod}, \ref{fig:Energies_Oblate_FreqMod}, and \ref{fig:Energies_Oblate_IntMod}. The fidelity function is better presented not in units of time but in cycles of excitation, given by $\frac{t}{T_{\mathrm{exc}}}$.

\begin{figure}[ht]
\includegraphics[width=\textwidth]{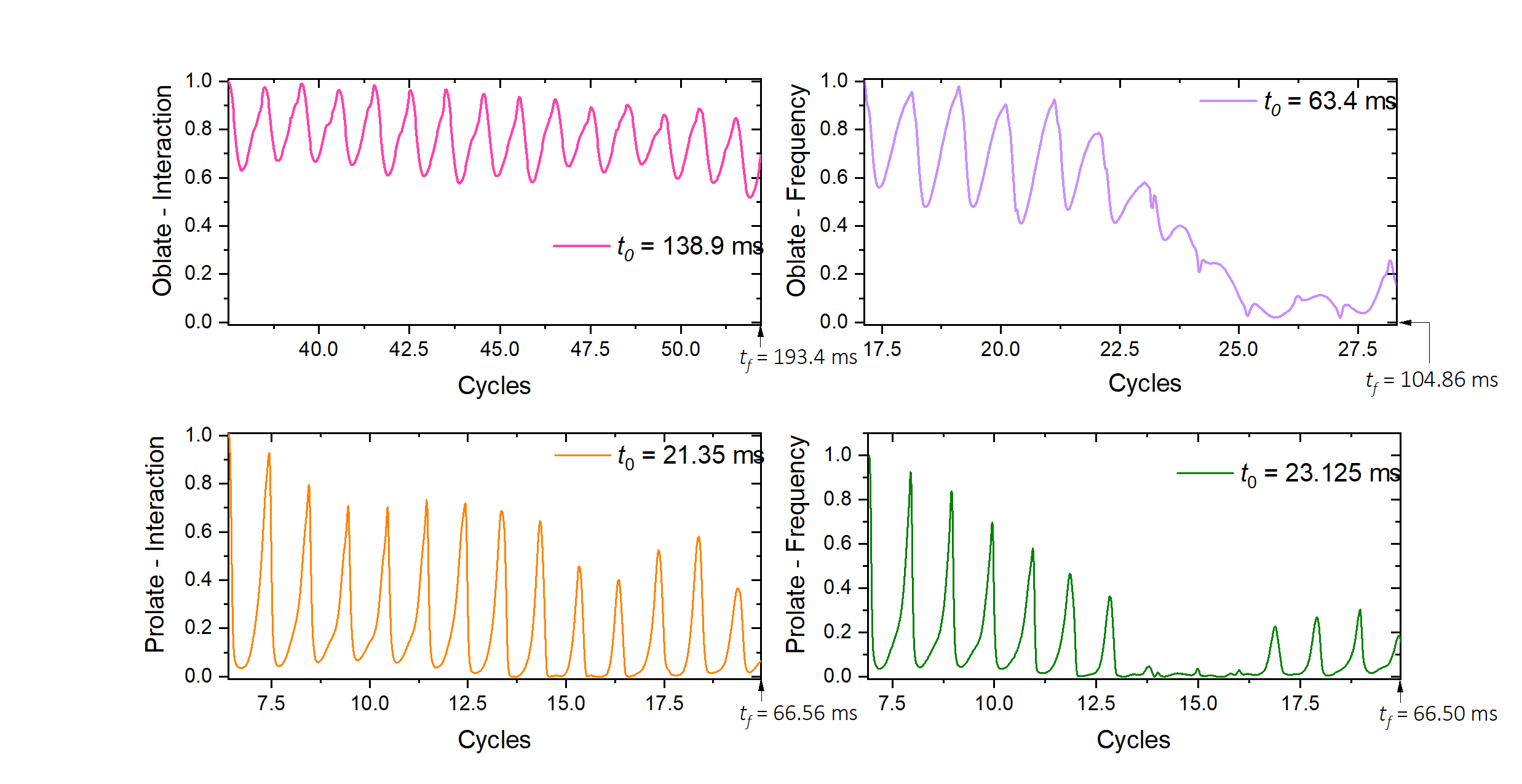}
\caption{\label{GFidelity_Prolate} Fidelity function for both parametric modulations for the oblate and prolate trap. The function \eqref{FidelityDef} is used to study the evolution of the formed density patterns. The initial wavefunction, $\Psi_{ab}(t)$, used to compare the wavefunction for further times, $\Psi_{ab}(t+dt)$, is selected so that the initial time $t$ coincides with the  increase of the kinetic energy. For the modulation through the scattering length it is observed that for both geometries, the patterns remain stable for a longer interval compared with the  behavior of the patterns when the modulation is introduced through one trap frequency. }
\end{figure}

 Thus, as shown in Figure \ref{GFidelity_Prolate}, we compare the evolution of this $xz$-plane wavefunction for the prolate trap in the seventh excitation cycle for the radial trap and scattering length modulations. For the oblate trap, the wavefunction at the 37th excitation cycle is taken for comparison when the modulation is through the axial trap frequency. For the scattering length modulation, the wavefunction for comparison is chosen in the 17th excitation cycle. Those times were chosen because they are close to the time at which the increasing of the kinetic energy is observed.

For the prolate trap, we observe that the system oscillates between an almost orthogonal state, since the inner product its zero, to a state that is very similar to the initial state. As we go far in time, the similitudes between both states decays. Nevertheless, the interval at which these oscillations occur is wide enough to be observed experimentally. 

The behavior between the two types of parametric modulation is different for the oblate trap. In this case, we observe that when the modulation is introduced through the scattering length, the system remains similar to the initial state. Thus, the fidelity function takes values above 0.5. For the modulation through the axial frequency, for a small interval of time, the system remains oscillating between similar states, which means that the wavefunction returns to have a very similar pattern in its density. After the 22nd excitation cycle, the system decays to another different state. Nevertheless, the ring patterns can be observed from the 17th - 22nd excitation cycles. 

Summarizing, for an oblate trap the fidelity function shows there is a difference between the axial frquency and scattering length modulations. This difference is observed in the stability and lifetime of the generated density patterns. Nevertheless, the behavior of both parametric modulations in the prolate trap is very similar. These observations match the results presented in  \cite{Shukuno-jphyssocjpn92} and complete their observation for an oblate trap.  

\begin{figure}[h]
\includegraphics[scale=0.5]{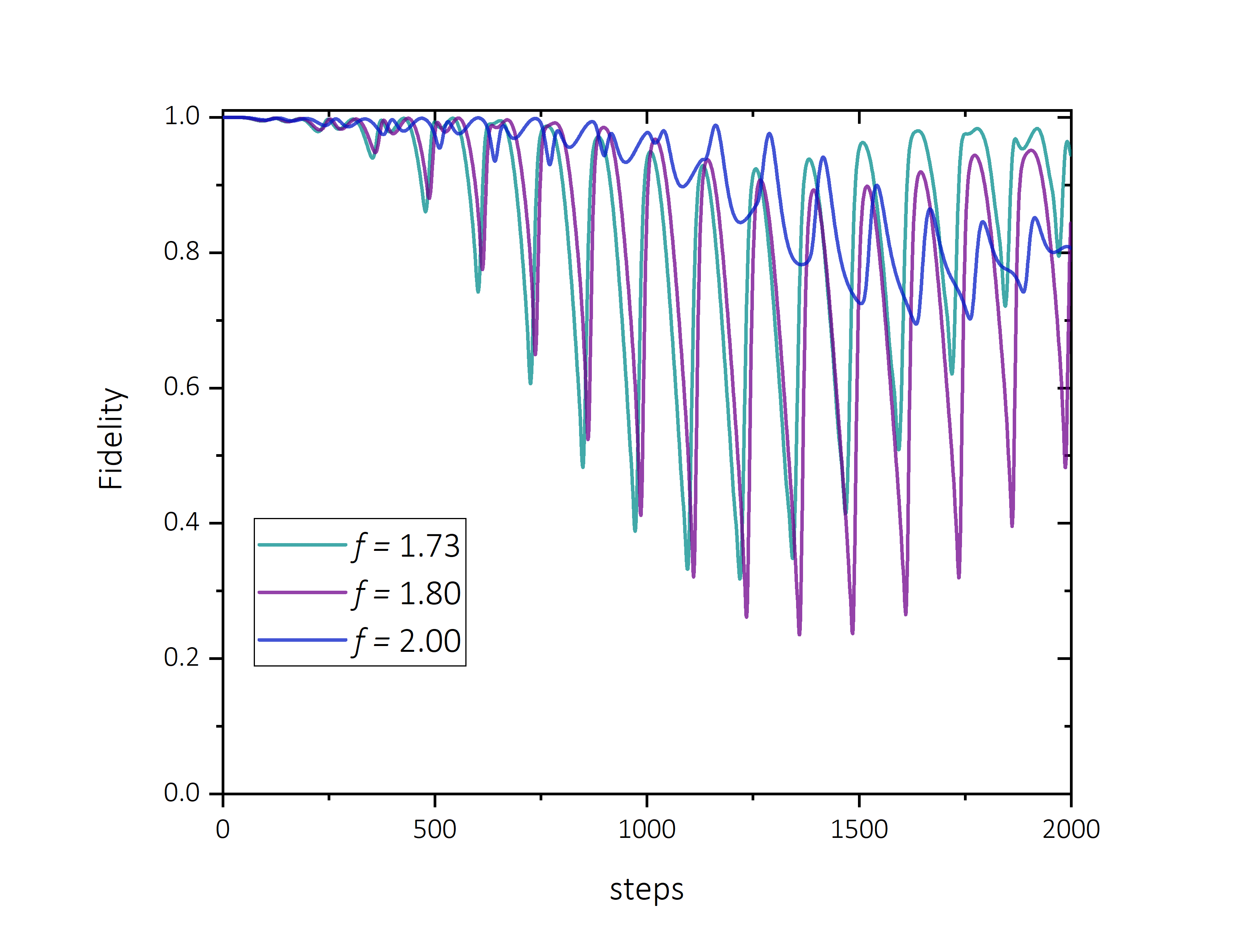}
\caption{\label{fig:comparison_fidelity} Fidelity comparison when the system is modulated with different frequencies. For an oblate trap with fixed frequencies $\omega_r^0=2\pi \times 10$ Hz and $\omega_z^ 0=2\pi \times 150$ Hz, the variational method states that the frequency associated with the breathing mode is given by $\Omega=f \omega_z$ with $f=1.80$.  If the system is modulated with a frequency around this value, the generated oscillations are smaller and decay faster. The $x$ axis shows the steps of the numerical simulation.  These steps are connected with time by $t=(\mathrm{steps}\times dt)\omega_z$, where $dt$, as defined in Section \ref{sec:numericalsimulation}, is $dt=\frac{T_{\mathrm{exc}}}{2^{13}}$ with $T_{\mathrm{exc}}=\frac{2\pi N_{\Omega}}{\Omega}$. }
\end{figure}

Finally, we use the fidelity function to compare the behavior of the system when it is modulated with different frequencies. As discussed previously in Section \ref{sec:parametricexcitation}, the frequency $\Omega$ at which the system is modulated affects the oscillations of the widths of the Gaussian wavefunction in the variational approach. These modifications are also reflected in the formation of the density patterns. In Figure \ref{fig:comparison_fidelity}, we use the Fidelity function to show that when an oblate trap is modulated with frequencies $\Omega=1.73\,\omega_z$ and $\Omega=2.0\,\omega_z$, the system responses oscillating since we are near the resonance frequency $1.80\, \omega_z$. However, these oscillations are smaller and decay faster. This theoretical expectations mean that patterns will appear sooner, and it would be more feasible to observe experimentally when the system is modulated near the resonance frequency. As we get far from the resonance frequency, patterns will take longer to appear. 

\section{Conclusions and Perspectives}\label{sec:Conclusions}

In this work we have presented a detailed study on the formation mechanism and dynamics of the spatial and temporal patterns that emerge in a driven BEC trapped in a 3D cylindrical symmetric harmonic potential. We consider two different trap geometries, either a prolate (cigar-shaped) system or an oblate (pancake-shaped) geometry.

For each geometry, we explore two different excitation mechanisms consisting in periodically modulating the most confining trap frequency or, alternatively, the scattering length of the system. The frequency of this modulation corresponds to the breathing mode frequency associated to the specific geometry of the condensate.

We start our study by computing the breathing mode frequency as a function of the geometry of the condensate employing a variational method. For very elongated geometries, we find that these frequencies are given by $\Omega = 1.8\omega_z$ and $\Omega = 2.0\omega_r$ for the oblate and prolate geometries, respectively.

Next, we investigate the dynamics of the driven systems using numerical simulations of the GPE. As a first result, we find that the trap geometry determines the type of the patterns that can be formed in the system. We observe fringe patterns for the prolate trap, and ring patterns for the oblate one. These patterns are resonantly formed when the system is driven at the corresponding breathing mode frequency. Small shifts from this resonance condition lead to short-lived patterns that take longer excitation times to emerge, as revealed by the fidelity function in Figure \ref{fig:comparison_fidelity}.

We have shown that the pattern formation can be understood in terms of the dynamics of the distribution of the energy in the system. Here, the kinetic energy of the condensate plays a special role. Initially, the energy that the excitation transfers into the system is employed to excite the breathing mode along the most confining direction of the condensate. After a certain time the kinetic energy increases also along the orthogonal less confined direction, onsetting the formation of either the fringe patterns in the prolate trap or the rings patterns in the oblate system. This indicates that the redistribution of the energy along the less confined direction plays a central role in the formation of the patterns, revealing the tridimensional nature of the observed phenomena. This last claim is supported by our calculations of the density current that shows that velocity field is relevant along both the most and least confined directions.

Another important result is the comparison between the two excitation mechanisms, in which the drive can be applied directly into the trap or into the interatomic interaction strength. Although the produced patterns with either mechanism are very similar, the time of appearance is shorter when modulating the trap. This is because the isotropy of the scattering length modulation stabilizes the system by dissipating the excitations more efficiently. This is of relevance on the design of experiments, in which the lifetime of the sample is a limiting factor.

 In addition, it is important to discuss that for the analysis of the obtained results in this work, a parametric excitation introduced through the least confined direction, i.e., axial frequency $\omega_z$ in the prolate trap and radial frequency $\omega_r$ for the oblate trap, was also considered to compare the answer of the system to the direction of excitation. The system response to this parametric modulation was observed to be very slow, preventing the formation of any density patterns at standard experimental observation times. 

Finally, the study of density patterns is still a very rich area with many properties to be studied. For example, in this work, it is found that a change in the geometry of the external potential modifies the time at which the energy redistribution, and thus the density patterns, appear. By changing the external potential, the system's density is modified, and consequently, the nonlinear term of the GPE equation. Nevertheless, the interaction between particles can also change the system's nonlinearity, and it would be interesting to understand how this variation is.  

Likewise, the observed patterns in this work are stable under an interval of time and imposed by an external potential. By changing the geometry of the external potential, it would be interesting to study if a pattern with $l \neq 0$ can arise.

\begin{acknowledgments}
This work was supported by DGAPA-UNAM-PAPIIT through grant number IN109021. 
Also, was supported by CONAHCyT through projects CF-2023-I-72 and A1-S-39242; DGAPA-UNAM-PAPIIT through grant numbers IN109021 and IN105724, and CIC-UNAM through projects LANMAC-2023 and LANAMC-2024. FJPC and AdRL acknowledge the support of the CONAHCYT project A1-S-39242. AdRL acknowledges the scholarship of the CONAHCYT. 
\end{acknowledgments}

\bibliographystyle{apsrev4-2}
\bibliography{bibringpatterns}

\end{document}